\documentclass[10pt,journal,a4paper]{IEEEtran} 

\usepackage{color}
\usepackage{cite}      
\usepackage[caption=false, font=footnotesize]{subfig}
\usepackage{graphicx}
\usepackage[latin1]{inputenc}
\usepackage{amsfonts}
\usepackage{amsmath}
\usepackage{times}
\usepackage{algpseudocode}
\usepackage{algorithm}
\usepackage{enumerate}
\usepackage{siunitx}
\usepackage{multirow}
\usepackage{tikz}

\usepackage{amssymb}

\newtheorem{lemma}{Lemma}[section]

\newtheorem{proposition}{Proposition}[section]

\newtheorem{example}{Example}[section]

\newcolumntype{M}[1]{>{\centering\arraybackslash}m{#1}}

\newcommand{\SSS}{{\mathrm{S}}}
\newcommand{\RRR}{{\mathrm{R}}}
\newcommand{\DDD}{{\mathrm{D}}}

\newcommand{\eSR}{{\epsilon_{\mathrm{SR}}}}
\newcommand{\teSR}{{\tilde{\epsilon}_{\mathrm{SR}}}}
\newcommand{\eSD}{{\epsilon_{\mathrm{SD}}}}
\newcommand{\eRD}{{\epsilon_{\mathrm{RD}}}}
\newcommand{\eSRj}{{\epsilon_{\mathrm{S} \mathrm{R}_j}}}
\newcommand{\eRjD}{{\epsilon_{\mathrm{R}_j \mathrm{D}}}}
\newcommand{\mRD}{{m_{\mathrm{RD}}}}

\newcommand{\FF}{{\mathbb{F}}}
\newcommand{\PP}{{\mathbb{P}}}
\newcommand{\CC}{{\mathbf{C}}}
\newcommand{\GG}{{\mathbf{G}}}
\newcommand{\rank}{{\mathrm{rank}}}

\newcommand{\gauss}{\genfrac[]{0pt}{}}

\begin{document}

\title{Reliability of Relay Networks under Random Linear Network Coding}

\author{Evgeny Tsimbalo\thanks{
E. Tsimbalo and M. Sandell are with The Telecommunications Research Laboratory of Toshiba Research Europe Ltd., Bristol, BS1 4ND, UK (email: \{E.Tsimbalo, Magnus.Sandell\}@toshiba-trel.com).}
 and Magnus Sandell}
\maketitle

\begin{abstract}
We consider a single-source, multiple-relay, single-destination lossy network employing Random Linear Network coding at all transmitting nodes. We address the problem of calculating the probability of successful decoding at the destination node. In contrast to some previous studies, we assume the classical RLNC scheme, in which the relaying nodes simply recode packets, without resorting to decoding. In addition, we consider an arbitrary field size and take into account correlation between the relay nodes. We propose a novel upper bound for an arbitrary number of relays, which becomes exact for a single relay. Using Monte Carlo simulations, we show that the proposed bound is very accurate, exhibiting the mean square error as low as $10^{-6}$. We also demonstrate the throughput gain of the proposed scheme over alternative coding and relaying strategies.
\end{abstract}


\section{Introduction\label{sec:Introduction}}

Relay-based topologies are widely used in modern wireless networks to extend transmission range and to utilise spatial diversity in order to improve communication reliability. For example, in body sensor networks, intermediate nodes are commonly deployed to relay data to monitoring devices \cite{Fafoutis2016}. Nevertheless, such networks are often characterised by poor reliability at the physical layer, resulting in frequent retransmissions requested by the upper layers. In this context, the idea of Application Layer Forward Error Correction (AL-FEC) \cite{Gomez-Barquero2009}, where the underlying layers are abstracted as a \emph{packet erasure channel} and coding is performed over packets rather than bits, has attracted significant interest over the years.

Traditional AL-FEC schemes are based on the so-called fountain codes \cite{Byers2002, Wang2016b}. However, fountain codes usually operate efficiently with large block sizes and are applied on a single-hop basis, \emph{i.e.}, requiring decoding followed by re-encoding at each hop \cite{Magli2013}. By contrast, the idea of linearly combining packets using random coefficients, known in the literature as \emph{Random Linear Network Coding} (RLNC) \cite{Ho2003, Ho2006}, dictates simple recoding at each hop, thus inherently supporting multihop topologies. It was proved that RLNC is capacity-achieving in the context of multicast wireless networks \cite{Lun2004, Lun2005}.

In networks operated under RLNC, the source message of $K$ packets is encoded by creating random linear combinations of the packets \cite{Lun2004}. To recover the source message, the traditional requirement is to collect $K$ linearly independent sets of coefficients at the destination node. This approach, also referred to as full-rank decoding, will be employed in this work. It should be noted that alternative, rank-deficient decoding techniques, whereby the source message can be recovered from less than $K$ linearly independent sets of coefficients, have also been considered in the literature \cite{7541795}.

Several efforts to analyse the performance RLNC applied to relay networks are present in the literature. Multiple-source, multiple-relay two-hop networks were studied in \cite{Chiasserini2013, Seong2014, Khan2016}. In \cite{Chiasserini2013}, each source node was designed to transmit a single uncoded packet, while each relay node encoded packets it received into a single linear combination and forwarded it to the destination node. An exact expression for the probability of successful decoding was derived, assuming that at most one of the source-relay links failed and that communication between the relays and the destination was error free. More general scenarios involving sparse RLNC \cite{Cooper2000} were considered in \cite{Seong2014}, for which lower and upper bounds were proposed. More accurate bounds were subsequently derived in \cite{Khan2016}. In both \cite{Seong2014, Khan2016} the degree of sparsity was determined by packet loss, which led to poor code performance in certain scenarios. Furthermore, it should be noted that in all three studies \cite{Chiasserini2013, Seong2014, Khan2016}, only a single packet was assumed to be transmitted by any source or relay node, and the number of necessary relay nodes was set to be equal to the number of required coded transmissions. As a result, a large number of relay nodes was required to provide reliability, which is often impractical.

Another common assumption in \cite{Chiasserini2013, Seong2014, Khan2016} was that the source nodes transmit uncoded packets. Encoding at the source node was  considered in \cite{Shi2013} for a single-source single-relay network with a direct link between the source and destination nodes. It was demonstrated that encoding at both source and relay nodes has performance gains compared to encoding at the relay only. The analysis, however, was performed assuming an infinitely large field size, which in practice is limited \cite{Pedersen2013, Ferreira2014}. A finite field size was considered in \cite{Khan2015}, where the relay node attempts the decoding of the source message, followed by recoding. Using the terminology of relay networks, such scheme can be referred to as \emph{Decode-and-Forward (DF)}. However, the analysis in \cite{Khan2015} was based on the assumption that the relay and destination nodes receive uncorrelated subsets of packets from the source. As a result, the derived performance bounds were accurate only under certain network scenarios. The correlation was taken into account in \cite{Tsimbalo2016a}, where an exact expression for the probability of successful decoding for a single-relay network operating under the DF scheme was derived. Furthermore, the DF scheme was complemented with a passive relay mode, in which if the relay is not able to decode, it simply re-transmits the packets it received from the source to the destination.

In this work, we address the limitations of the previous studies of relay networks employing RLNC and provide the following contributions:
\begin{itemize}
\item We generalise the single-relay scenario considered previously in \cite{Khan2015, Tsimbalo2016a} to an arbitrary number of relays. In contrast to \cite{Chiasserini2013, Seong2014, Khan2016}, we enable encoding at the source node to further mitigate packet loss. In addition, we generalise the approach of \cite{Chiasserini2013, Seong2014, Khan2016} by allowing each relay node to transmit \emph{multiple} coded packets, thus resulting in a fewer relays to achieve reliable communication. We propose a tight upper bound for the probability of successful decoding, which is exact in the case of a single relay. We also propose an approximation of the bound, which is accurate for networks with a sufficiently large number of relays. 
\item Compared with the DF scheme employed in \cite{Khan2015, Tsimbalo2016a}, we consider a different relaying strategy, whereby the relay nodes do not attempt to decode, but simply recode the packets they receive from the source. Once again employing the terminology of relay networks, we call such scheme \emph{Recode-and-Forward (RF)}. The RF strategy follows the original idea of RLNC, in which every transmitting node in the network performs random linear combinations of the packets in its buffer. Comparing with DF, the RF scheme is expected to provide better reliability, since the relay nodes are likely to transmit more packets.
\item We perform thorough benchmarking of the proposed analytical results via extensive Monte Carlo simulation for various network and code parameters. In particular, we demonstrate that the proposed bound is exact for a single relay and is tight for an arbitrary number of relays.
\item We demonstrate the benefit of RLNC-based coding compared with an uncoded scenario and a repetition code in terms of throughput. In addition, we benchmark the proposed RF relaying strategy with the DF scheme, as well as with a case of no coding at the source node.
\end{itemize}

The remainder of the paper is organised as follows. Section~\ref{sec:System-model} describes the system model and provides some background results. Section~\ref{sec:Theory} presents theoretical analysis, starting with a general case of an arbitrary relay, followed by some special cases. In Section~\ref{sec:NumResults}, the proposed results are compared with simulated ones and a throughput analysis is presented. The conclusions are drawn in Section~\ref{sec:Conclusions}.

\section{System Model and Problem Formulation\label{sec:System-model}}

\begin{figure}
\begin{centering}
\includegraphics{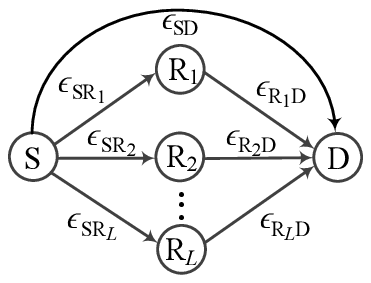}
\par\end{centering}
\caption{Block diagram of a two-relay network.\label{fig:drawing-relay-network}}
\end{figure}

A single-source network with $L$ relays is depicted in Fig.~\ref{fig:drawing-relay-network}. The goal is to successfully deliver a message of $K$ equally-sized source packets from the source node $\SSS$ to the destination node $\DDD$ with the help of the relay nodes $\RRR_j$, $j=1,\ldots,L$. Let $\eSD$, $\eSR_j$, and $\eRjD$ denote packet erasure probabilities (PEP) of links  $\SSS \rightarrow \DDD$, $\SSS \rightarrow \RRR_j$, and $\RRR_j \rightarrow \DDD$, respectively. 

In the RF scheme considered in this work, the transmission is performed in two stages. During the first stage, the source node broadcasts $N_\SSS \geq K$ coded packets. For encoding, we employ non-systematic RLNC, traditionally used in multihop networks due to its simplicity and scalability \cite{Magli2013}. Each coded packet is a linear combination of the source packets with the coefficients drawn uniformly at random from a finite field $\FF_q$ of size $q$. Let $m_j \leq N_\SSS$, $j=1,\ldots,L$, denote the number of coded packets received by $\RRR_j$, and let $m_\DDD$ denote the number of coded packets received by $\DDD$ from $\SSS$. It is assumed that each receiving node knows the coding coefficients associated with every packet it receives, which can be achieved by transmitting the  coefficients in the packet header \cite{Ho2006}. Each relay can therefore construct an $m_j \times K$ matrix of coefficients $\CC_{\SSS \rightarrow \RRR_j}$, $j=1,\ldots,L$, while the destination node can reconstruct an $m_\DDD \times K$ matrix of coefficients $\CC_{\SSS \rightarrow \DDD}$.

During the second stage of the RF scheme, each relay node recodes the packets it received from $\SSS$ into $N_\RRR$ packets and transmits them to $\DDD$. For recoding, the $j$-th relay node generates a new $N_\RRR \times m_j$ matrix of random coefficients $\GG_j$ and applies it to its received packets. Since the code is linear, this is equivalent to encoding the original, $K$ source packets by a new, recoded set of coefficients arranged in an $N_\RRR \times K$ matrix $\GG_j \CC_{\SSS \rightarrow \RRR_j}$, $j=1,\ldots,L$. The new recoded coefficients are transmitted from each relay in the packet headers. 

Let $m'_j \leq N_\RRR$ denote the number of packets the destination node receives from $\RRR_j$, $j=1,\ldots,L$. Node $\DDD$ can  therefore reconstruct $m'_j$ rows of matrix $\GG_j \CC_{\SSS \rightarrow \RRR_j}$. After stacking the reconstructed rows from all relays together, in addition to the rows $\DDD$ recovered directly from $\SSS$, a total matrix of coding coefficients $\CC_{\DDD}$ reconstructed by $\DDD$ is obtained: 
\begin{equation}
	\CC_{\DDD} =  \begin{bmatrix} \CC_{\SSS \rightarrow \DDD} \\
						         \CC_{\RRR \rightarrow \DDD} \\						   
					     \end{bmatrix},\label{eq:dest_mat}
\end{equation}
where
\begin{equation}
	\CC_{\RRR \rightarrow \DDD} = \mathrm{diag} \left( \tilde{\GG}_1, \ldots, \tilde{\GG}_L \right) \begin{bmatrix} \CC^T_{\SSS \rightarrow \RRR_1}  & \hspace{-2mm}\ldots\hspace{-2mm} & \CC^T_{\SSS \rightarrow \RRR_L} \end{bmatrix}^T \label{eq:recoded_coeffs}
\end{equation}
is an $m' \times K$ matrix of all recoded coefficients received from the relay nodes, $m'=\sum_{j=1}^L m'_j$, and $\tilde{\GG}_j$ denotes an $m'_j \times m_j$ matrix of recoding coefficients obtained from $\GG_j$ by removing rows corresponding to packets lost between $\RRR_j$ and $\DDD$. The destination node can recover the source message if matrix $\CC_{\DDD}$ \eqref{eq:dest_mat} has rank $K$. Our goal will be to characterise the performance of such network in terms of \emph{probability of successful decoding}, which will be also referred to as \emph{decoding probability}.

\subsection{Theoretical Background\label{sec:Background}}
We now present some relevant results in the area of RLNC that we will use throughout the paper.

Consider first a single-hop point-to-point link with PEP $\epsilon$. The decoding probability of a message of $K$ packets after $N$ coded transmissions is expressed as follows \cite{Khan2015}:
\begin{equation}
	P(N, \epsilon) = \sum_{m = K}^N \mathcal{B}(m, N, \epsilon) \PP(m,K).\label{eq:prob_dec_ptp}
\end{equation}
Here, $\mathcal{B}(m, N, \epsilon)$ denotes the probability mass function (PMF) of the binomial distribution, \emph{i.e.}, the probability of $m$ successes out of $N$ independent trials with success probability $(1-\epsilon)$, calculated as follows:
\begin{equation}
	\mathcal{B}(m, N, \epsilon) = \binom{N}{m} (1 - \epsilon)^m \epsilon^{N - m}.\label{eq:binom}
\end{equation}
Furthermore, $\PP(m,K)$ in \eqref{eq:prob_dec_ptp} denotes the probability that an $m \times K$ matrix of elements generated uniformly at random from $\mathbb{F}_q$, $m \geq K$, has rank $K$, which is given by \cite{Trullols-Cruces2011}:
\begin{equation}
	\PP(m,K) = \prod_{i=0}^{K-1}(1-q^{i-m}).\label{eq:prob_full_rank}
\end{equation}
It should be noted that \eqref{eq:prob_full_rank} is a special case of the probability that an $m \times K$ matrix of elements generated uniformly at random from $\mathbb{F}_q$ has rank $r \leq \min(m,K)$ given by\cite{Khan2018}:
\begin{equation}
	{\PP}_r (m, K) = \frac{1}{q^{m(K-r)}} \gauss{K}{r}_{q} \prod_{i=0}^{r-1} \left(1-q^{i-m}\right),\label{eq:prob_rank_r}
\end{equation}
where
\begin{equation}
	\gauss{K}{r}_{q} = \prod_{i=0}^{r-1} \frac{q^K-q^{i}}{q^{r}-q^{i}} = \prod_{i=0}^{r-1} \frac{q^{K-i}-1}{q^{r-i}-1} \label{eq:gaussian}
\end{equation}
is the Gaussian binomial coefficient \cite{VanLint}. 

Structured matrices of elements generated uniformly at random from $\mathbb{F}_q$ have been also studied the literature. In particular, it was shown in \cite{Tsimbalo2016a} that for a block matrix $\mathbf{X}$ formed by two vertically stacked matrices $\mathbf{A}$ and $\mathbf{B}$ with dimensions $a \times K$ and $b \times K$, the probability that $\mathbf{X}$ has rank $K$, given $a+b \geq K$, is given by
\begin{equation}
	\mathrm{Pr}\left[ \rank(\mathbf{X}) = K \right] = \sum_{i=\max(0,K-b)}^{\min(a,K)} \PP_i(a,K) \PP(b, K-i). \label{eq:prob_rank_vert}
\end{equation}
Based on \eqref{eq:prob_rank_vert}, it was shown in \cite{Tsimbalo2016a} that the joint probability of two random matrices $\mathbf{X}_1$ and $\mathbf{X}_2$ with dimensions $m_1 \times K$ and $m_2 \times K$, $m_1,m_2 \geq K$, and $m_{12}$ common rows having simultaneously rank $K$ is given by
\begin{equation}
	\PP^{(2)}(\mathbf{m}, K) = \sum_{i} \PP_{i}(m_{12},K) \prod_{j=1}^2 \PP(m_{j}-m_{12},K-i),\label{eq:prob_two_mat_fr}
\end{equation}
where $\mathbf{m}'=(m_1,m_2,m_{12})$ and the summation is performed over the values of $i$ from $\max(0,K-m_{1}+m_{12},K-m_{2}+m_{12})$ to $\min(m_{12},K)$.

\begin{table}
\begin{centering}
\caption{Notation used throughout the paper.}\label{tab:Notation}
\ifCLASSOPTIONtwocolumn
\begin{tabular}{|c|m{6.5cm}|}
\else
\begin{tabular}{|c|m{10cm}|}
\fi
\hline 
Notation & Description\tabularnewline
\hline 
\hline
$K$ & Number of packets forming an information message\\ 
\hline
$N_\SSS$ & Number of coded packet transmissions from $\SSS$\\ 
\hline
$N_\RRR$ & Number of coded packet transmissions from each $\RRR$\\ 
\hline
$q$ & Size of the finite field under consideration\\ 
\hline
$L$ & Number of relay nodes\\ 
\hline
$\eSD$ & PEP between $\SSS$ and $\DDD$\\ 
\hline
$\eSR_j$ & PEP between $\SSS$ and $\RRR_j$\\ 
\hline
$\eRjD$ & PEP between $\RRR_j$ and $\DDD$ \\ 
\hline
$m_j$ & Number of packets received by $\RRR_j$ from $\SSS$\\ 
\hline
$m'_j$ & Number of packets received by $\DDD$ from $\RRR_j$\\ 
\hline
$\CC_{\SSS \rightarrow \RRR_j}$ & $m_j \times K$ matrix of coding coefficients received by $\RRR_j$ from $\SSS$  \\ 
\hline
$\CC_{\SSS \rightarrow \DDD}$ & $m_\DDD \times K$ matrix of coding coefficients received by $\DDD$ directly from $\SSS$\\ 
\hline
$\GG_j$ & $N_\RRR \times m_j$ matrix of recoding coefficients generated by $\RRR_j$\\ 
\hline
$\CC_{\RRR \rightarrow \DDD}$ & $m' \times K$ matrix of recoded coefficients received by $\DDD$ from all relays \\  
\hline
$\CC_{\DDD}$ & $m' + m_\DDD \times K$ matrix of all coding coefficients collected by $\DDD$\\ 
\hline 
$P(N, \epsilon)$ & Decoding probability for a point-to-point link with $N$ coded transmissions and PEP $\epsilon$\\
\hline
$\mathcal{B}(m, N, \epsilon)$ & PMF of the binomial distribution: the probability of $m$ successes out of $N$ trials with success probability $\epsilon$\\
\hline 
$\PP(m,K)$ & Probability that an $m \times K$ matrix of elements generated uniformly at random from $\mathbb{F}_q$ is full rank\\
\hline 
$\PP_r(m,K)$ & Probability that an $m \times K$ matrix of elements generated uniformly at random from $\mathbb{F}_q$ has rank $r \leq \min(m,K)$\\
\hline 
$\PP^{(2)}(\mathbf{m}, K)$ & Joint probability that two matrices with common rows have rank $K$\\
\hline
\end{tabular}
\par\end{centering}
\end{table}

The notation introduced in this section is summarised in Table~\ref{tab:Notation}.

\section{Theoretical Analysis of Decoding Probability\label{sec:Theory}}

During the first transmission stage in the relay network described in Section~\ref{sec:System-model}, the source node multicasts to the relay and destination nodes. Each transmitted packet can be received by a single relay, a selection of at least two relays or by none of the relays. As a result of receiving common packets, the matrices of coding coefficients $\{\CC_{\SSS \rightarrow \RRR_j}\}^{L}_{j=1}$ reconstructed by the relay nodes may have shared rows and are therefore correlated \cite{Tsimbalo2018}. In total, there are $\sum_{i=1}^{L}\binom{L}{i}=2^{L}-1$ groups of packets corresponding to all possible combinations of at least one relay.

Let $\CC_{\SSS \rightarrow \RRR}$ denote a matrix constructed from the union of all rows of $\{\CC_{\SSS \rightarrow \RRR_j}\}^{L}_{j=1}$, such that each row occurs in $\CC_{\SSS \rightarrow \RRR}$ only once. Let $m$ denote the number of rows of $\CC_{\SSS \rightarrow \RRR}$. For the $i$-th row of $\CC_{\SSS \rightarrow \RRR}$, $i=1,\ldots,m$, let the indices of the relays that simultaneously received the corresponding packet be denoted as $J_i \subseteq \{1,\ldots,L\}$. We now construct an equivalent $m' \times m$ recoding matrix $\tilde{\GG}$ such that matrix $\CC_{\RRR \rightarrow \DDD}$ \eqref{eq:recoded_coeffs} can be expressed as follows:
\begin{equation}
	\CC_{\RRR \rightarrow \DDD} = \tilde{\GG} \CC_{\SSS \rightarrow \RRR}. \label{eq:recoded_coeffs3}
\end{equation}
To this end, the $m'_j$ elements of $\tilde{\GG}$ at the intersection of the $j$-th group of rows, corresponding to the $j$-th relay, $j=1,\ldots,L$, and the $i$-th column, $i=1,\ldots,m$, are taken from the column of $\tilde{\GG}_j$ corresponding to the $i$-th packet if $j \in J_i$, or set to zeros otherwise. In other words, $\CC_{\SSS \rightarrow \RRR}$ in \eqref{eq:recoded_coeffs3} can be thought of as a global matrix of unique coding coefficients received by all relays, while $\tilde{\GG}$ can be thought of as a global matrix of recoding coefficients. It should be noted that no assumptions are made on whether any given relay node is aware of common packets it shares with other relays, and \eqref{eq:recoded_coeffs3} is just a more compact, equivalent representation of \eqref{eq:recoded_coeffs}.

\begin{example}\label{example:G_matrix}
Let $L=3$. Without loss of generality, let the rows of coefficients received by a subset of relays be grouped together. For instance, matrix $\CC_{\SSS \rightarrow \RRR}$ can be written in the following form:
\begin{equation}
	\CC_{\SSS \rightarrow \RRR} = \begin{bmatrix}\CC^T_{123} & \CC^T_{12} & \CC^T_{13} & \CC^T_{23} & \CC^T_{11} & \CC^T_{22} & \CC^T_{33} \end{bmatrix}^T, \nonumber
\end{equation}
where ${\CC}_{123}$ contains the rows of coefficients common to all three relays, and ${\CC}_{jk}$, $j,k = 1,2,3$, contains the rows shared between the $j$-th and $k$-th relays. In total, matrix $\CC_{\SSS \rightarrow \RRR}$ consists of $2^L-1 = 7$ vertical blocks. The corresponding equivalent matrix $\tilde{\GG}$ has the following structure:
\begin{equation}
	\tilde{\GG} = \begin{bmatrix} \tilde{\GG}^{(1)}_{123} & \tilde{\GG}_{12}   & \tilde{\GG}_{13}   & \mathbf{0}   & \tilde{\GG}_{11}     & \mathbf{0}   & \mathbf{0} \\
					   			  \tilde{\GG}^{(2)}_{123} & \tilde{\GG}_{21}   & \mathbf{0} & \tilde{\GG}_{23}     & \mathbf{0}   & \tilde{\GG}_{22}     & \mathbf{0} \\
								  \tilde{\GG}^{(3)}_{123} & \mathbf{0} & \tilde{\GG}_{31}   & \tilde{\GG}_{32}     & \mathbf{0}   & \mathbf{0}   & \tilde{\GG}_{33}   \\
					     \end{bmatrix}.\nonumber
\end{equation}
Here, matrices $\tilde{\GG}^{(j)}_{123}$, $j=1,2,3$, are composed of the columns of $\tilde{\GG}_j$ that correspond to the packets received by all three relays, while matrices $\tilde{\GG}_{jk}$, $j,k=1,2,3$, are composed of the columns of $\tilde{\GG}_j$ that correspond to packets shared between the $j$-th and $k$-th relays. The zero matrices correspond to packets not received by at most two relays. To see that matrix $\tilde{\GG} \CC_{\SSS \rightarrow \RRR}$ contains all rows of $\CC_{\RRR \rightarrow \DDD}$ given by \eqref{eq:recoded_coeffs}, consider the following numerical example. Let $K=4$ and assume that each relay received two packets from the source and reconstructed the following matrices of coding coefficients:
\begin{equation}
	\CC_{\SSS \rightarrow \RRR_1} = \begin{bmatrix} 1 & 0 & 0 & 1 \\
													0 & 1 & 0 & 1 \\	
	\end{bmatrix}, \CC_{\SSS \rightarrow \RRR_1} = \begin{bmatrix} 0 & 1 & 0 & 1 \\
																   1 & 1 & 1 & 0 \\	
												   \end{bmatrix},\nonumber
\end{equation}
\begin{equation}
	\CC_{\SSS \rightarrow \RRR_3} = \begin{bmatrix} 1 & 1 & 1 & 0 \\
													1 & 1 & 1 & 1 \\	
	\end{bmatrix},\nonumber
\end{equation}
where one packet is shared by $\RRR_1$ and $\RRR_2$, and another is shared by $\RRR_2$ and $\RRR_3$. Assume also that the destination node received two packets from each relay and recovered the following matrices of recoding coefficients:
\begin{equation}
	\tilde{\GG}_1 = \begin{bmatrix} 1 & 0 \\
									1 & 1 \\	
	\end{bmatrix}, \tilde{\GG}_2 = \begin{bmatrix} 0 & 1 \\
												   1 & 0 \\	
								   \end{bmatrix}, \tilde{\GG}_3 = \begin{bmatrix} 1 & 1 \\
																			      1 & 0 \\	
								   								  \end{bmatrix}.\nonumber
\end{equation}
Matrix $\CC_{\SSS \rightarrow \RRR}$ can be constructed from the union of all rows of $\{\CC_{\SSS \rightarrow \RRR_j}\}^{3}_{j=1}$ as follows:
\begin{equation}
	\CC_{\SSS \rightarrow \RRR} = \begin{bmatrix} \CC_{12} \\
							                    \CC_{23} \\ 
							                    \CC_{11} \\
							                    \CC_{33} \\
							      \end{bmatrix} = \begin{bmatrix} 0 & 1 & 0 & 1 \\
												  1 & 1 & 1 & 0 \\
												  1 & 0 & 0 & 1 \\
												  1 & 1 & 1 & 1 \\	 
							                    \end{bmatrix}, \nonumber
\end{equation}
which corresponds to the following matrix $\tilde{\GG}$:
\begin{equation}
	\tilde{\GG} = \begin{bmatrix} \tilde{\GG}_{12}   & \mathbf{0}   		& \tilde{\GG}_{11}  & \mathbf{0} \\
					   			  \tilde{\GG}_{21}   & \tilde{\GG}_{23}     & \mathbf{0}   		& \mathbf{0} \\
								  \mathbf{0} 		 & \tilde{\GG}_{32}     & \mathbf{0}   		& \tilde{\GG}_{33}   \\
					     \end{bmatrix} = \begin{bmatrix} 0 & 0 & 1 & 0 \\
					                                     1 & 0 & 1 & 0 \\
					                                     0 & 1 & 0 & 0 \\
					                                     1 & 0 & 0 & 0 \\
					                                     0 & 1 & 0 & 1 \\
					                                     0 & 1 & 0 & 0 \\
					                     \end{bmatrix}.\nonumber
\end{equation}
Here, the first column of $\tilde{\GG}$, for instance, corresponds to the packet shared by $\RRR_1$ and $\RRR_2$: its first two elements form the second column of $\tilde{\GG}_1$, the second two elements form the first column of $\tilde{\GG}_2$ and the last two elements are set to zero. It can be seen that matrix $\tilde{\GG} \CC_{\SSS \rightarrow \RRR}$ contains all rows of coefficients received by the destination node determined by \eqref{eq:recoded_coeffs}.
\end{example}
\vspace{\baselineskip}

As described in Section~\ref{sec:System-model}, the destination node can recover the source message if matrix $\CC_{\DDD}$ of all coding coefficients reconstructed by the destination node, given by \eqref{eq:dest_mat}, has rank $K$. Let $X$ define such event:
\begin{equation}
	X: \rank(\CC_{\DDD}) = K. \label{eq:event}
\end{equation}
We now address the problem of finding the probability of $X$.

Employing \eqref{eq:recoded_coeffs3}, matrix $\CC_{\DDD}$ can be expressed as follows:
\begin{equation}
	\CC_{\DDD} =  \begin{bmatrix} \mathbf{I}_\DDD & \mathbf{0}\\
				                            \mathbf{0} & \tilde{\GG} \end{bmatrix} \begin{bmatrix} \CC_{\SSS \rightarrow \DDD} \\
						   																   \CC_{\SSS \rightarrow \RRR} \end{bmatrix},\label{eq:dest_mat2}
\end{equation}
where $\mathbf{I}_\DDD$ is an $m_\DDD \times m_\DDD$ identity matrix. By analogy to the correlation effect between the relays, a packet received by a subset of relays can also be received by $\DDD$. Let $\mathcal{M}_{\RRR \DDD} \subseteq \{1,\ldots,m\}$ denote a subset of rows of $\CC_{\SSS \rightarrow \RRR}$ corresponding to such packets, and let $\mRD = |\mathcal{M}_{\RRR \DDD}|$ denote the number of these packets. Using this notation, it can be observed that the columns of $\tilde{\GG}$ with indices from $\mathcal{M}_{\RRR \DDD}$ create linear combinations of coding coefficients already known by $\DDD$, hence the rank of $\CC_{\DDD}$ does not depend on the elements in these columns. Let the remaining columns of $\tilde{\GG}$, corresponding to packets with indices \mbox{$\{1,\ldots,m\} \setminus \mathcal{M}_{\RRR \DDD}$}, form an $m' \times (m-\mRD)$ matrix $\hat{\GG}$, and let $r$ denote its rank. From \eqref{eq:dest_mat2}, it can be observed that given $r$, matrix $\CC_{\DDD}$ will have $r+m_\DDD$ independently generated rows. Therefore, employing \eqref{eq:prob_full_rank} and marginalising over the distribution of $r$, the probability that $\CC_{\DDD}$ has rank $K$ can be calculated as follows:
\begin{equation}
	\mathrm{Pr}\left[ X \right] = \sum_{r = r_{\min}}^{r_{\max}} \mathrm{Pr}\left[ \rank(\hat{\GG}) = r \right] \PP(r+m_\DDD, K), \label{eq:prob_dest_full_rank}
\end{equation}
where \mbox{$r_{\min} = \max (0,K-m_\DDD)$} is chosen such that \mbox{$r + m_\DDD \geq K$} and \mbox{$r_{\max} = \min \left( m', m - \mRD \right)$}.

From \eqref{eq:prob_dest_full_rank}, it is clear that the probability distribution of the rank of matrix $\hat{\GG}$ needs to be calculated. As can be observed from Example~\ref{example:G_matrix}, even for $L=3$, matrix $\tilde{\GG}$, and hence matrix $\hat{\GG}$, has a complex structure. Even if the rank distribution was calculated exactly, enumerating all possible combinations of numbers of received packets and marginalising over their distributions would produce a result too complex for practical use.

Instead of finding an exact expression for the probability that $\hat{\GG}$ has a certain rank, we establish the following bound:
\begin{lemma}\label{lemma:prob_block_rank_r_bound}
The probability that the \mbox{$m' \times (m-\mRD)$} matrix $\hat{\GG}$ has a certain rank $r$  is upper-bounded as follows:
\begin{equation}
	\mathrm{Pr}\left[ \rank(\hat{\GG}) = r \right] \leq \PP_r \left( m', m - \mRD \right),\label{eq:lemma_bound}
\end{equation}
where $\PP_r \left( m', m-\mRD \right)$ is the probability that an $m' \times (m-\mRD)$ matrix of random coefficients has rank $r$, which is given by \eqref{eq:prob_rank_r}.
\end{lemma}
%
\begin{IEEEproof}
It can be observed that bound \eqref{eq:lemma_bound} is based on replacing deterministic zero elements in $\hat{\GG}$ arising from the packet loss between the source and relay nodes with random elements generated from $\FF_q$. Intuitively, allowing those elements to be non-zero increases the probability of the matrix to have the same rank, hence the bound is an upper one. The equality in \eqref{eq:lemma_bound} will correspond to the case when all relays share the same set of received packets; in other words, when there are no deterministic zero elements in $\hat{\GG}$.
\end{IEEEproof}

Bound \eqref{eq:lemma_bound} is expected to be tighter for lower PEP values between the source and relay nodes. Indeed, in such case a packet transmitted by $\SSS$ is more likely to be received by all relays, thus making the number of zero elements in $\hat{\GG}$ smaller. Furthermore, for the specific case of a single-relay network, \eqref{eq:lemma_bound} turns into an equality.

Based on \eqref{eq:lemma_bound}, we now establish a bound for the probability \eqref{eq:prob_dest_full_rank} that matrix $\CC_{\DDD}$ has rank $K$:

\begin{lemma}\label{lemma:prob_dest_full_rank}
The probability of event $X$ that the matrix of coding coefficients $\CC_{\DDD}$ at the destination node has rank $K$  is upper-bounded as follows:
\begin{equation}
	\mathrm{Pr}\left[ X \right] \leq \PP^{(2)} \left( \mathbf{m}' ,K \right),\label{eq:lemma2_bound}
\end{equation}
where $\mathbf{m}'=(m'+m_\DDD,m-\mRD+m_\DDD,m_\DDD)$ and the right-hand side is a joint probability that two matrices with dimensions an $m'+m_\DDD \times K$ and $m-\mRD+m_\DDD \times K$ with $m_\DDD$ common rows are simultaneously full rank, which is given by \eqref{eq:prob_two_mat_fr}. 
\end{lemma}
%
\begin{IEEEproof}
See Appendix~A.
\end{IEEEproof}

Having obtained the upper bound for the probability that the matrix of coding coefficients at the destination node has rank $K$ for given $m$, $m_\DDD$, $\mRD$ and \mbox{$m' = \sum_{j=1}^{L} m'_j$}, we now formulate the bound for the overall decoding probability of the $L$-relay network:
%
\begin{proposition}\label{proposition:prob_dec_L}
The decoding probability $P_\RRR^{(L)}$ of an $L$-relay network operated under the RF scheme is upper-bounded as follows:
\setlength{\arraycolsep}{0.14em} 
\begin{eqnarray}\hspace{-1em}
	P_\RRR^{(L)} & \leq & \sum_{\mathbf{m}} \alpha\left( \mathbf{m}, N_\SSS, \boldsymbol{\epsilon} \right) \sum_{m'_1,\ldots,m'_L} \prod_{j=1}^{L} \mathcal{B}(m'_j, N_\RRR, \eRjD) \nonumber\\	       
	       &   & \hspace{38mm}\cdot \PP^{(2)} \left( \mathbf{m}', K \right), \label{eq:prob_dec_L}
\end{eqnarray}
\setlength{\arraycolsep}{5pt}%
where 
\setlength{\arraycolsep}{0.14em}
\begin{eqnarray}
	\alpha\left( \mathbf{m}, N_\SSS, \boldsymbol{\epsilon} \right) & = & \binom{N_\SSS}{\mRD} \hspace{-1mm} \binom{N_\SSS-\mRD}{m-\mRD} \hspace{-1mm} \binom{N_\SSS-m}{m_\DDD-\mRD} \nonumber \\
	&  & \cdot (1-\teSR)^{m} \teSR^{N_\SSS-m} \nonumber \\
	&  & \cdot (1-\eSD)^{m_\DDD} \eSD^{N_\SSS-m_\DDD},\label{eq:multinom}
\end{eqnarray}
\setlength{\arraycolsep}{5pt}%
\mbox{$\mathbf{m} = (m,m_\DDD,\mRD), \boldsymbol{\epsilon} = (\teSR,\eSD)$}, \mbox{$\teSR = \prod_{j=1}^L \eSR_j$} and the summation is performed over the following values:
\setlength{\arraycolsep}{0.14em} 
\begin{eqnarray}
	m + m_\DDD & \geq & K; m,m_\DDD \leq N_\SSS; \nonumber\\
	\mRD & \geq & \max(0,m+m_\DDD-N_\SSS), \mRD \leq \min(m,\mRD); \nonumber\\
	m' &\geq& K-m_\DDD; m'_j \leq N_\RRR, j=1,\ldots,L. \label{eq:prob_dec_L_indices}
\end{eqnarray}
\setlength{\arraycolsep}{5pt}%

\end{proposition}
%
\begin{IEEEproof}
Consider the upper bound \eqref{eq:lemma2_bound}. It can be observed that this probability depends on several random variables: $m$, the total number of unique packets received by all $L$ relays, $m_\DDD$, the number of packets received by $\DDD$ from $\SSS$, $\mRD$, the number of packets commonly received by at least a single $\RRR$ and $\DDD$, and $m'$, the total number of packets received by $\DDD$ from all relays. Consider the first three variables - $m$, $m_\DDD$ and $\mRD$ - describing possible numbers of packets received from $\SSS$ at the first transmission stage. As was shown in \cite{Tsimbalo2016a, Tsimbalo2018}, their joint PMF is that of the multinomial distribution, which describes the probability of a particular combination of numbers of occurrences of mutually exclusive outcomes out of $N_\SSS$ trials and is given by \eqref{eq:multinom}. Here, the collection of $L$ source-to-relay links is considered as a single link with an equivalent PEP \mbox{$\teSR = \prod_{j=1}^L \eSR_j$}. On the other hand, the total number of packets received at the second stage, $m'$, is a sum of independent random variables $m'_j$, $j=1,\ldots,L$, each described by the binomial distribution $\mathcal{B}(m'_j, N_\RRR, \eRjD)$. By marginalising the right-hand side of \eqref{eq:lemma2_bound} over the distribution of all variables in question, \eqref{eq:prob_dec_L} can be readily obtained. The starting values of $m$, $m_\DDD$ and $m'_j$ are chosen such that at least $K$ packets are collected by $\DDD$ in total. The starting value of $\mRD$ is chosen such that it equals $0$ if $m + m_\DDD \leq N_\SSS$ and $m + m_\DDD - N_\SSS$ otherwise, to avoid unnecessary summation terms.
\end{IEEEproof}

We now consider some special cases of the $L$-relay network described in Section~\ref{sec:System-model}.

\subsection{Single-relay Network\label{sec:Theory_Single}} 

When the network has only a single relay ($L=1$), matrix $\tilde{\GG}$, defined by \eqref{eq:recoded_coeffs3}, and hence matrix $\hat{\GG}$, obtained from $\tilde{\GG}$ by removing $\mRD$ columns, does not have any deterministic zeros. As a result, bounds \eqref{eq:lemma_bound} and \eqref{eq:lemma2_bound}, and therefore the proposed bound \eqref{eq:prob_dec_L}, become exact. The decoding probability can be calculated as follows:
\begin{equation}\hspace{-1em}
	P_\RRR^{(L)} = \sum_{\mathbf{m}} \alpha\left( \mathbf{m}, N_\SSS, \boldsymbol{\epsilon} \right) \sum_{m'} \mathcal{B}(m', N_\RRR, \eRD) \PP^{(2)} \left( \mathbf{m}', K \right) \hspace{-1mm}. \label{eq:prob_dec_1}
\end{equation}
Comparing \eqref{eq:prob_dec_1} with \eqref{eq:prob_dec_L}, it can be seen that the bound for the general case of an arbitrary $L$ approximates the $L$-relay network as a single-relay one, with an equivalent source-to-relay link having the PEP equal to \mbox{$\teSR = \prod_{j=1}^L \eSR_j$}. During the second transmission stage, the equivalent relay performs $L$ rounds of $N_\RRR$ coded transmissions with the PEP $\eRjD$ in each round, $j=1,\ldots,L$.

\subsection{Multiple-relay Network, $\eSD = 1$\label{sec:Theory_noSD}}
In the case when there is no communication between $\SSS$ and $\DDD$, \emph{i.e.}, $\eSD = 1$, the destination node does not receive any packets at the first transmission stage. In other words, \mbox{$m_\DDD = \mRD = 0$}. As a result, the bound of Lemma~\ref{lemma:prob_dest_full_rank} turns into the probability of two uncorrelated matrices with $m'$ and $m$ rows having rank $K$:
\begin{equation}
	\Pr[X] \leq \PP(m',K) \PP(m,K). \label{eq:lemma2_bound_noSD}
\end{equation}
In addition, the multinomial distribution \eqref{eq:multinom} becomes binomial. Consequently, $\eqref{eq:prob_dec_L}$ can be rewritten as follows:
\setlength{\arraycolsep}{0.14em} 
\begin{eqnarray}\hspace{-1em}
	P_\RRR^{(L)} & \leq & \sum_{m=K}^{N_\SSS} \mathcal{B}(m, N_\SSS, \boldsymbol{\epsilon}) \sum_{m'_1,\ldots,m'_L} \prod_{j=1}^{L} \mathcal{B}(m'_j, N_\RRR, \eRjD) \nonumber\\	       
	       &   & \hspace{38mm}\cdot \PP(m',K) \PP(m,K). \label{eq:prob_dec_L_noSD}
\end{eqnarray}
\setlength{\arraycolsep}{5pt}%
Noting that $\PP(m,K)$ does not depend on $m'_j$ and employing \eqref{eq:prob_dec_ptp} yields
\setlength{\arraycolsep}{0.14em} 
\begin{eqnarray}\hspace{-1em}
	P_\RRR^{(L)} & \leq & P(N_\SSS, \teSR) \nonumber\\	       
	       &   & \cdot \sum_{m'_1,\ldots,m'_L} \prod_{j=1}^{L} \mathcal{B}(m'_j, N_\RRR, \eRjD) \PP(m',K). \label{eq:prob_dec_L_noSD2}
\end{eqnarray}
\setlength{\arraycolsep}{5pt}%
It can be observed that bound \eqref{eq:prob_dec_L_noSD2} consists of two independent parts. Indeed, the first term in \eqref{eq:prob_dec_L_noSD2} is the decoding probability of a point-to-point link he PEP equal to \mbox{$\teSR = \prod_{j=1}^L \eSR_j$}. The remainder of \eqref{eq:prob_dec_L_noSD2} is the decoding probability of another point-to-point link, in which $L$ rounds of $N_\RRR$ coded transmissions with the PEP $\eRjD$ in each round are performed, $j=1,\ldots,L$. In other words, by analogy to the general case of $\eSD \leq 1$, bound \eqref{eq:prob_dec_L_noSD2} treats the multiple-relay network as a single-relay equivalent. In the special case of a single relay, bound \eqref{eq:prob_dec_L_noSD2} becomes exact, as in the case of $\eSD \leq 1$:
\begin{equation}
	P_\RRR^{(1)} = P(N_\SSS, \eSR) P(N_\RRR, \eRD).\label{eq:prob_dec_single}
\end{equation}
In other words, the decoding probability of a single-relay network is the product of those for each individual link (hop)\footnote{Naturally, the result can be generalised to networks with any number of hops, also known as multiple-link tandem networks \cite{Lun2004}}.

\subsection{Approximation of Bound \eqref{eq:prob_dec_L}\label{sec:Theory2}}

From the general expression for the bound of the decoding probability \eqref{eq:prob_dec_L}, it can be observed that it contains multiple nested summations, the number of which grows with $L$, the number of relays in the network. As a result, bound \eqref{eq:prob_dec_L} is only of practical use when the number of relays is relatively small. In this section, we propose a simple approximation of the bound, the complexity of which does not depend on the network size.

\begin{figure}[t]
\includegraphics[width=1\columnwidth]{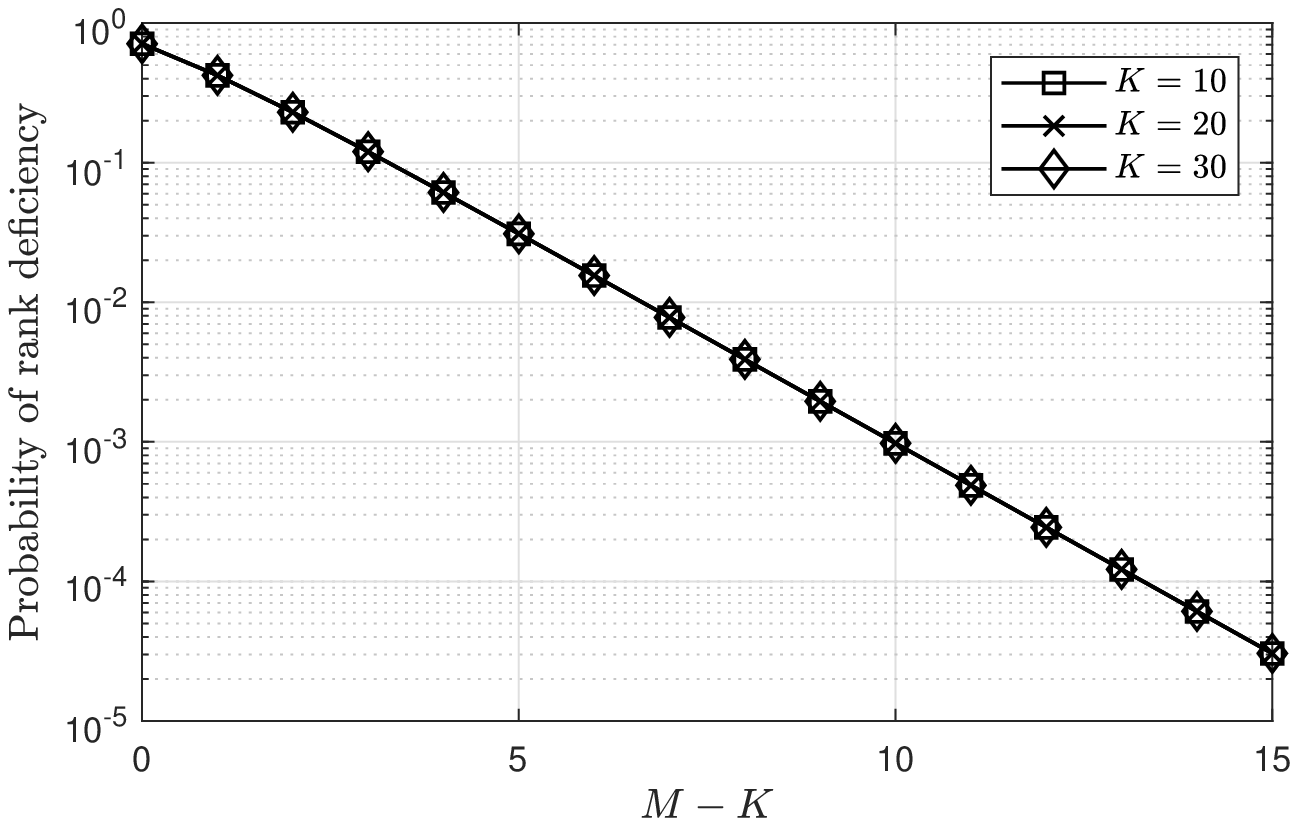}
\caption{Probability of an $M \times K$, $M \geq K$, binary matrix being rank deficient (\emph{i.e.}, having rank smaller than $K$), as a function of $M-K$.}
\label{fig:probRankDeficiency}
\end{figure}

Consider an $M \times K$ matrix of elements generated uniformly at random from $\FF_q$, $M \geq K$. It is well known \cite{Trullols-Cruces2011} that the probability of this matrix having full rank, given by \eqref{eq:prob_full_rank}, depends largely on $M-K$, rather than on $K$. In addition, for a sufficiently large $M-K$, this probability is close to one. Indeed, consider Fig.~\ref{fig:probRankDeficiency} illustrating the probability that the matrix is rank-deficient, which is equal to \mbox{$1-\PP(M,K)$}, as a function of $M-K$ for different values of $K$. It can be observed that $\PP(M,K)$ can be closely approximated to $1$ when $M-K \geq 15$, with the approximation error not larger than $2 \cdot 10^{-5}$ regardless of $K$.

Consider now the last term in \eqref{eq:prob_dec_L}, the joint probability that two matrices with dimensions an $(m'+m_\DDD) \times K$ and $(m-\mRD+m_\DDD) \times K$ with $m_\DDD$ common rows are simultaneously full rank. The first matrix has a number of rows equal to the total number of packets received by the destination node. The expected value of this number can be calculated as follows:
\begin{equation}
	\mathbb{E}[m'+m_\DDD] = K \left[ 1-\eSD + L (1-\eRD) \right] / R,\label{eq:expectation}
\end{equation}
where for simplicity it is assumed that \mbox{$\eRjD =\eRD$} and \mbox{$N=N_\SSS=N_\RRR$}, and 
\begin{equation}
	R = K / N \label{eq:code_rate}
\end{equation}
denotes the code rate. The $(m'+m_\DDD) \times K$ matrix is expected to be full-rank with a high probability if the following condition holds:
\begin{equation}
	K \left(\left[ 1-\eSD + L (1-\eRD) \right] / R - 1 \right) \geq \delta, \label{eq:condition}
\end{equation}
where, based on Fig.~\ref{fig:probRankDeficiency} and the discussion earlier, $\delta$ can be set to $15$. 

\begin{example}
Consider a network with $K=20$, $R=0.8$, $\eSD=1$ and $\eRD = 0.5$. Condition \eqref{eq:condition} holds for $\delta=15$ if the number of relays $L \geq 3$.
\end{example}

We now formulate a simplified version of bound \eqref{eq:prob_dec_L}.

\begin{proposition}[Simplified bound]
The decoding probability of an $L$-relay network is upper-bounded by the simplified version of bound \eqref{eq:prob_dec_L} as follows:
\begin{equation}
	P_\RRR^{(L)} \leq P(N_\SSS, \tilde{\epsilon}),\label{eq:simple_bound}
\end{equation}
where $P(N_\SSS, \tilde{\epsilon})$ is the decoding probability of a point-to-point link with PEP 
\begin{equation}
	\tilde{\epsilon} = \eSD \prod_{j=1}^L \eSRj.\label{eq:equivPER}
\end{equation}
\end{proposition}
\begin{IEEEproof}
Consider the last term in \eqref{eq:prob_dec_L}, the joint probability of two correlated matrices being full rank. If condition \eqref{eq:condition} holds for a sufficiently large $\delta$, the first of the two matrices, the one with dimensions $(m'+m_\DDD) \times K$, is full rank with a high probability, resulting in the following tight bound:
\begin{equation}
	\PP^{(2)} \left( \mathbf{m}', K \right) \leq \PP(m-\mRD+m_\DDD, K).\label{eq:prop2_proof_bound}
\end{equation}
It can be observed that bound \eqref{eq:prop2_proof_bound} depends only on \mbox{$m-\mRD+m_\DDD$}, the total number of unique packets received by the destination node and all relays nodes. Let $\tilde{m}$ denote this number. By viewing the destination node and all relay nodes as a single equivalent node receiving packets from the source, $\tilde{m}$ can be thought of as a number of successes out of $N_\SSS$ trials with success probability \mbox{$(1-\tilde{\epsilon})$}, where $\tilde{\epsilon}$ is given by \eqref{eq:equivPER}. In other words, $\tilde{m}$ is distributed binomially with the PMF equal to $\mathcal{B}(\tilde{m}, N_\SSS, \tilde{\epsilon})$. By marginalising bound \eqref{eq:prop2_proof_bound} and employing \eqref{eq:prob_dec_ptp}, \eqref{eq:simple_bound} can be readily obtained.
\end{IEEEproof}

It can be observed that the proposed simplified bound does not depend on the PEP of the relay-to-destination links, assuming that there are enough recoded transmissions during the second stage to provide the required number of linearly independent sets of coding coefficients to $\DDD$. Under this assumption, to achieve successful decoding, it is sufficient to collect $K$ linearly independent sets of coding coefficients during the first stage. The new approximation is tight to the original bound \eqref{eq:prob_dec_L} as long as condition \eqref{eq:condition} is satisfied. From \eqref{eq:condition}, it can be seen that the simplified bound becomes tighter as the number of relays $L$ increases, which makes it a suitable alternative to \eqref{eq:prob_dec_L} for larger networks, as will be demonstrated in Section~\ref{sec:NumResults}.

Approximating the decoding probability of the network in question to that of an equivalent point-to-point link allows various RLNC analysis tools, existing in the literature in the context of point-to-point communication, to be applied to relay networks. For instance, the expected number of transmissions required to deliver a message over a point-to-point link was studied in \cite{Chatzigeorgiou2017} by viewing the decoding probability as the Cumulative Distribution Function (CDF) of the number of overhead packets, $N_\SSS - K$. Other performance metrics, such as throughput and coding gain, can also be derived from the decoding probability. Finally, the simplicity and scalability of bound \eqref{eq:simple_bound} enables comparison with other relaying strategies without resorting to extensive simulations, which will be illustrated in Section~\ref{sec:NumResults}.

\section{Numerical Results\label{sec:NumResults}}
In this section, we analyse the performance of the $L$-relay network described in Section~\ref{sec:System-model} via Monte Carlo simulation and compare the results with those predicted by the derived analytical framework. For Monte Carlo simulation, we employed the Kodo C++ network coding library \cite{Pedersen2011}. Each simulated result was obtained by averaging over $10^5$ iterations. For simplicity, it was assumed that all $\SSS \rightarrow \RRR$ links have the same PEP \mbox{$\eSR = \eSR_j$}, and all $\RRR \rightarrow \DDD$ links have the same PEP \mbox{$\eRD = \eRjD$},  $j=1,\ldots,L$. In addition, each node was assumed to transmit the same number of coded packets \mbox{$N = N_\SSS = N_\RRR$}.

\begin{figure}[t]
\includegraphics[width=1\columnwidth]{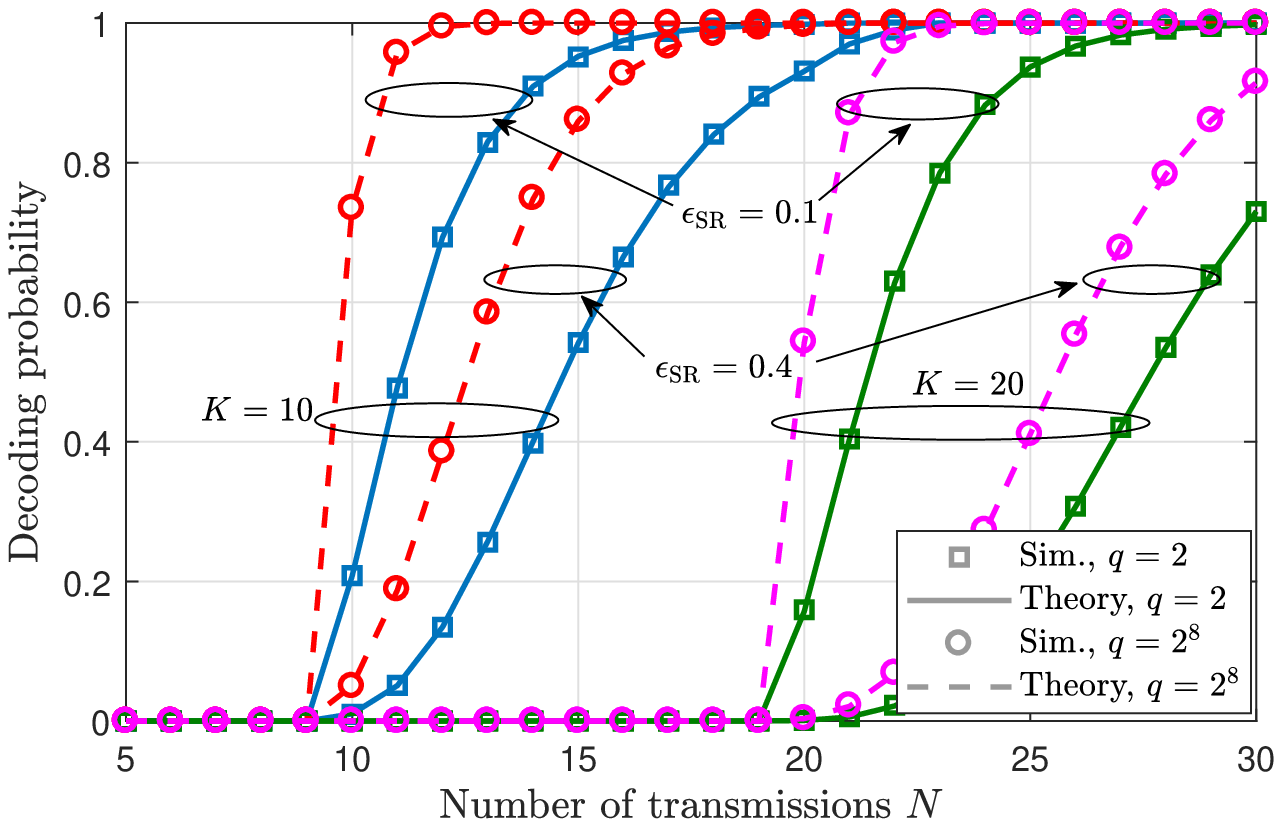}
\caption{Simulated and theoretical decoding probability as predicted by \eqref{eq:prob_dec_1} for a single relay, $K\in \left\lbrace 10,20\right\rbrace$, $q\in \left\lbrace 2,2^8\right\rbrace$, $\eSR=\eRD \in  \left\lbrace 0.1,0.4\right\rbrace$ and $\eSD = \eSR + 0.2$.}
\label{fig:res_nR1}
\end{figure}

We start with a single-relay network and compare the results predicted by the exact expression \eqref{eq:prob_dec_1} with simulated ones. Fig.~\ref{fig:res_nR1} illustrates the decoding probability as a function of the number of transmissions $N$ for various combinations of the source message size $K$, finite field size $q$ and PERs $\eSR$, $\eRD$ and $\eSD$. It can be observed that the performance predicted by the exact expression \eqref{eq:prob_dec_1} perfectly matches the simulated results in all cases.

\begin{figure}[t]
\subfloat[$L=2$]{\label{fig:res_nR2}
	\includegraphics[width=1\columnwidth]{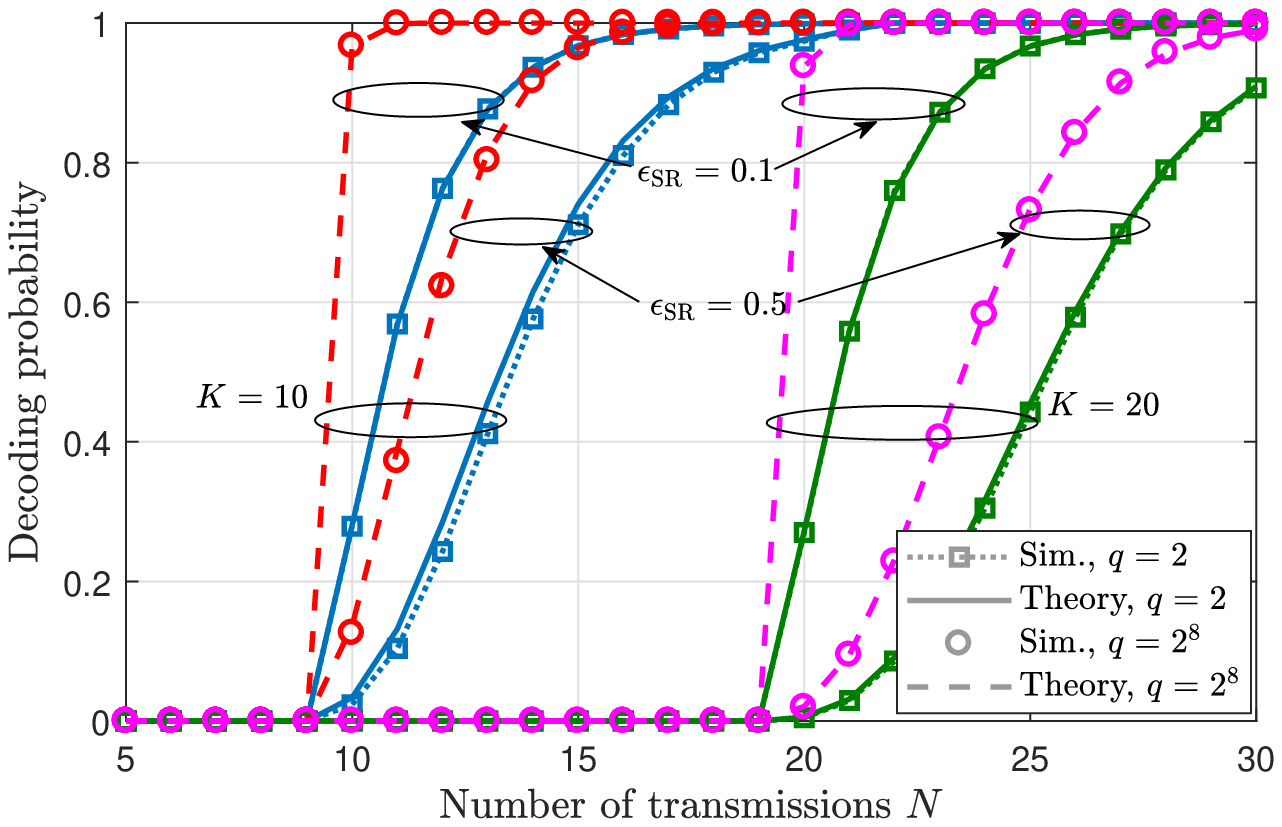}
}\\
\subfloat[$L=3$]{\label{fig:res_nR3}
	\includegraphics[width=1\columnwidth]{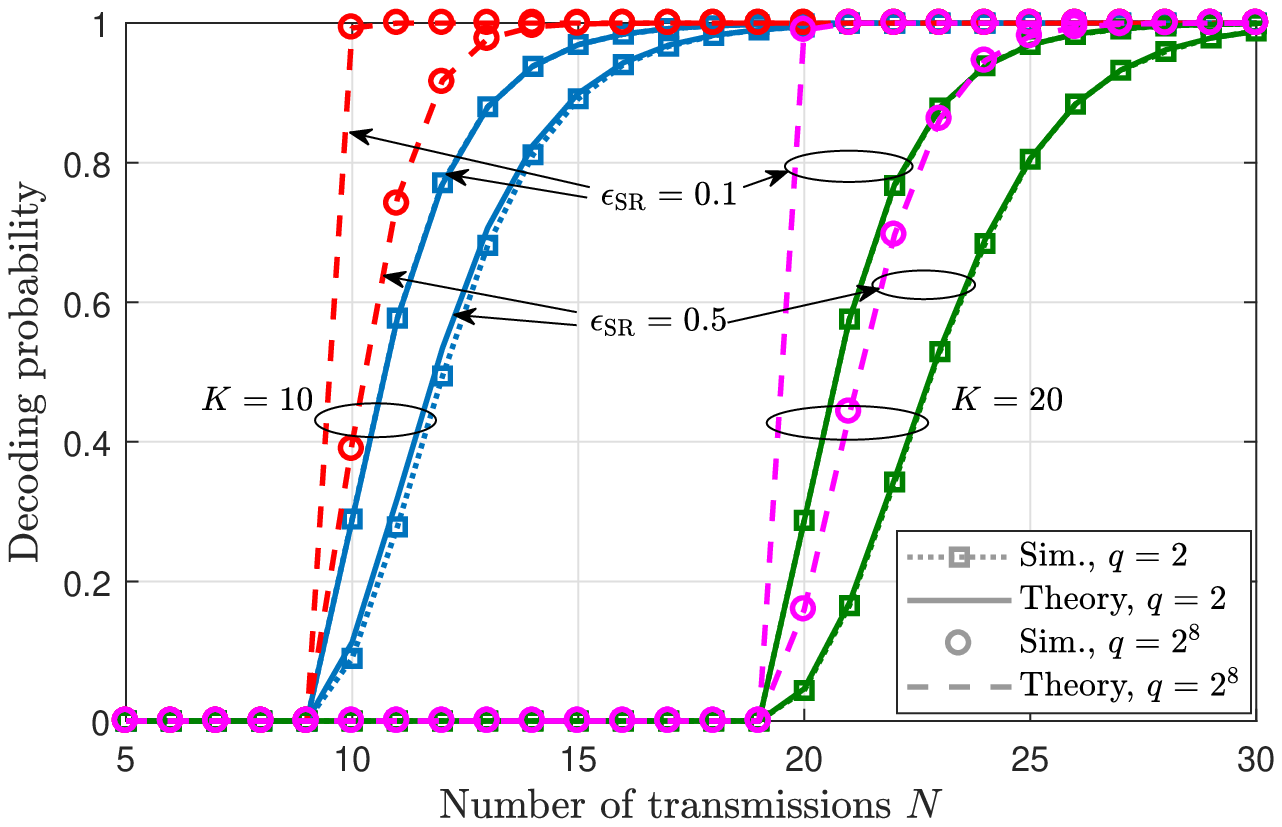}
}\\
\caption{Simulated and theoretical decoding probability as predicted by bound \eqref{eq:prob_dec_L} for two (a) and three (b) relays, $K\in \left\lbrace 10,20\right\rbrace$, $q\in \left\lbrace 2,2^8\right\rbrace$, \mbox{$\eSR=\eRD \in  \left\lbrace 0.1,0.5\right\rbrace$} and $\eSD = \eSR + 0.2$.}
\label{fig:res_nR23}
\end{figure}

We now consider two- and three-relay networks and employ the upper bound \eqref{eq:prob_dec_L} to predict the theoretical results, as illustrated in Fig.~\ref{fig:res_nR23}. It can be observed that bound \eqref{eq:prob_dec_L} accurately predicts the performance in the case of $\eSR=\eRD=0.1$ for both finite field sizes, and in the case of $\eSR=\eRD=0.5$ for $q=2^8$. The reason is that for the smaller $\eSR$, the relays are likely to share more packets, thus making approximation \eqref{eq:lemma_bound} closer. When the field size is large, the accuracy of bound \eqref{eq:lemma_bound} is high regardless of $\eSR$, because matrix $\hat{\GG}$ is likely to be full rank. On the other hand, a slight gap can be seen in the case of $\eSR=\eRD=0.5$ and the binary field. Again, this is explained by smaller correlation between the relays as $\eSR$ increases, which results in a larger number of deterministic zeros in matrix $\hat{\GG}$ and looser bound \eqref{eq:lemma_bound}. The gap, however, appears to be smaller for the larger $K$ and $L$. 

\begin{figure}[t]
\includegraphics[width=1\columnwidth]{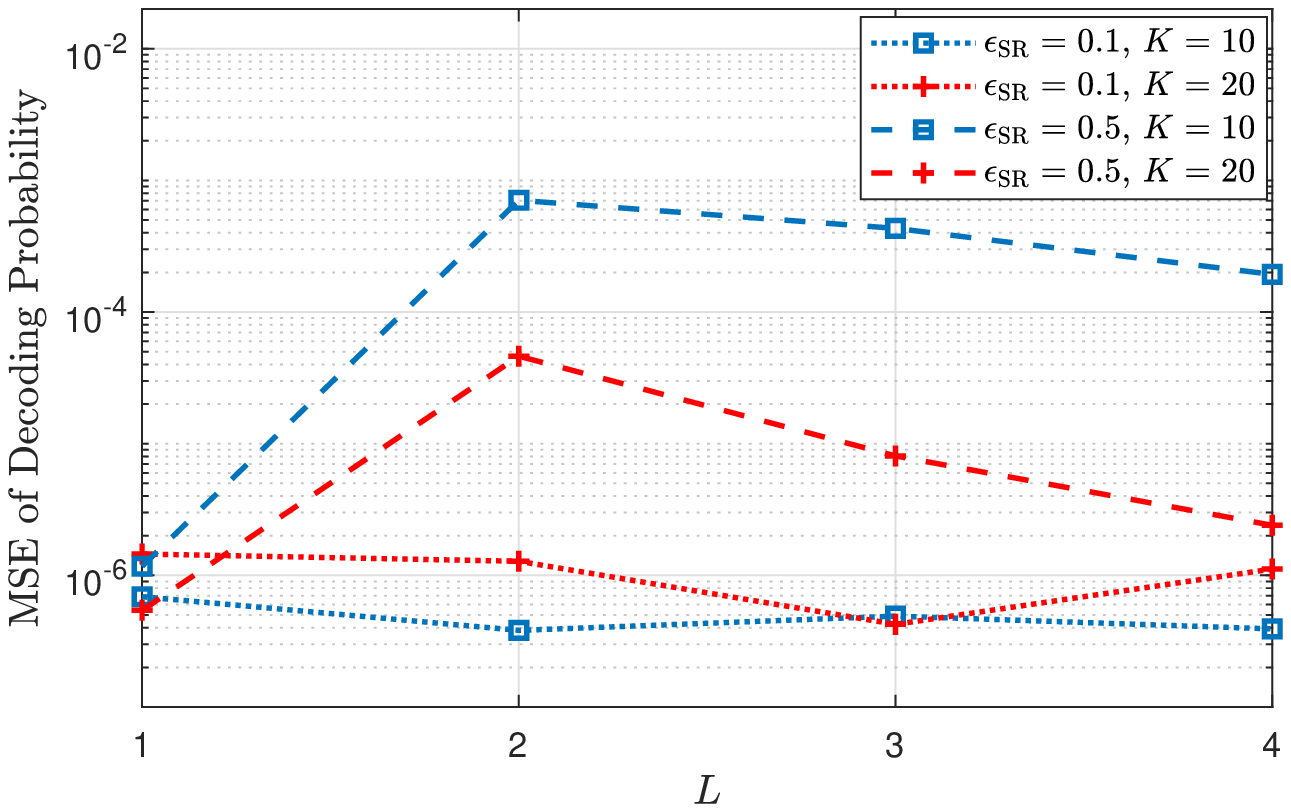}
\caption{MSE between the simulated and theoretical decoding probability as predicted by bound \eqref{eq:prob_dec_L} for $K\in \left\lbrace 10,20\right\rbrace$, $q=2$, \mbox{$\eSR=\eRD \in  \left\lbrace 0.1,0.5\right\rbrace$} and $\eSD = \eSR + 0.2$.}
\label{fig:res_MSE_q2}
\end{figure}

To illustrate the point further, Fig.~\ref{fig:res_MSE_q2} plots the Mean Squared Error (MSE) between bound \eqref{eq:prob_dec_L} and simulated results as a function of the number of relays $L$, for different numbers of relays and different values of $\eSR$ and $K$. The MSE was obtained by averaging the squared absolute difference between bound \eqref{eq:prob_dec_L} and simulated results over all possible numbers of transmissions from $N = K$ to $K+10$. For $L=1$, the MSE is of the order of $10^{-6}$ and is due to the averaging error of the Monte Carlo method. The MSE stays at this level as the number of relays increases, provided that $\eSR=0.1$. In contrast, for $\eSR=0.5$, the MSE increases sharply for $L=2$ and then gradually reduces as the number of relays grows. To understand this phenomenon, consider again matrix $\hat{\GG}$. The expected number of its rows and columns is equal to \mbox{$L N (1 - \eRD)$} and \mbox{$N \eSD(1 - \epsilon_{\SSS \RRR}^L)$}, respectively. Clearly, the difference between the two increases with $L$, and for a sufficiently large $L$ matrix $\hat{\GG}$ is full rank with a high probability, despite the presence of fixed zeros due to packet loss. In other words, both sides of inequality \eqref{eq:lemma_bound} will get closer to $1$ as $L$ increases, thus making bound \eqref{eq:prob_dec_L} tighter. For the same reason, the accuracy of the bound increases with~$K$. 

\begin{figure}[t]
\includegraphics[width=1\columnwidth]{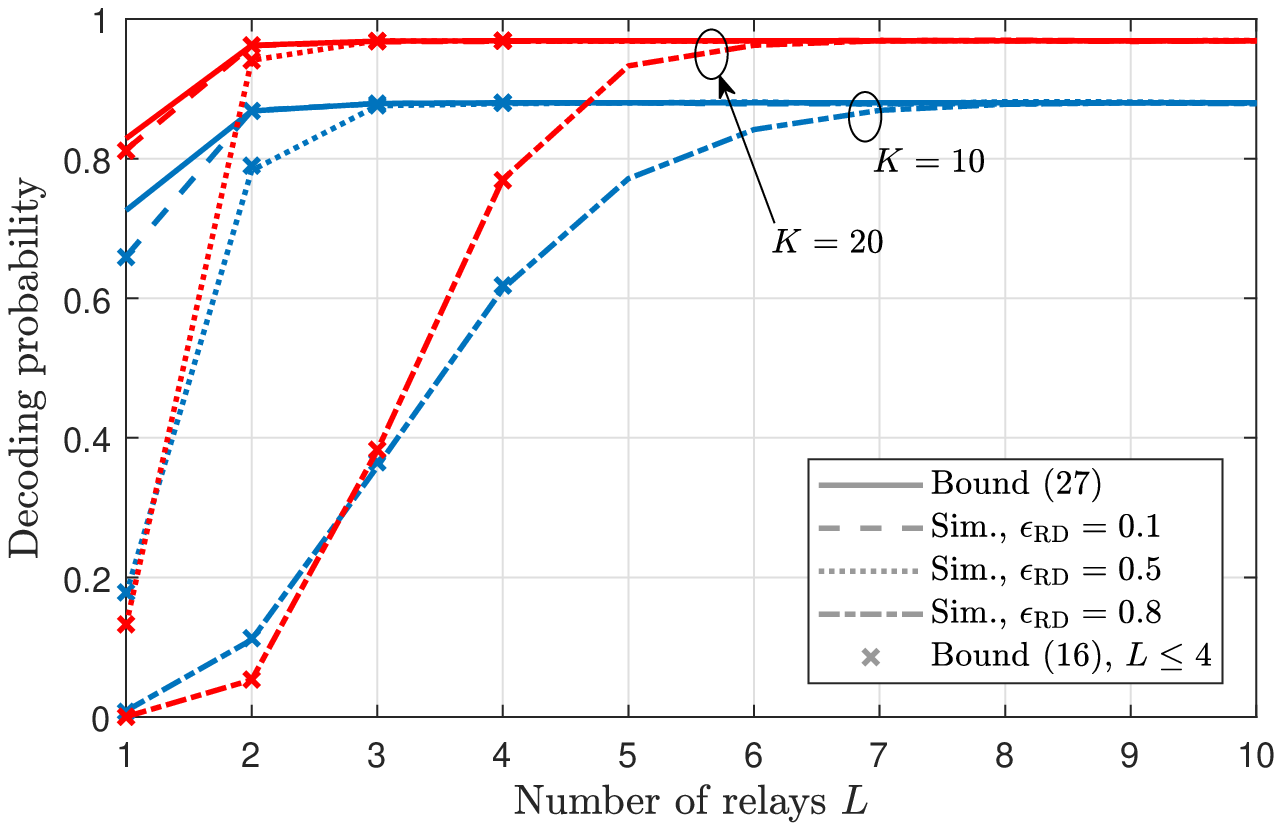}
\caption{Comparison of the decoding probability as predicted by bound \eqref{eq:simple_bound} with simulated results and bound \eqref{eq:prob_dec_L} for $K\in \{ 10,20\}$, $R=0.8$, $q=2$, $\eSR=0.1$, $\eSD=0.8$ and various $\eRD \in \{ 0.1,0.5,0.8 \}$.}
\label{fig:res_Simple_bound}
\end{figure}

We now investigate the accuracy of bound \eqref{eq:simple_bound}, a simplified version of bound \eqref{eq:prob_dec_L}. Fig.~\ref{fig:res_Simple_bound} illustrates the comparison between bound \eqref{eq:simple_bound} and simulated results for up to $10$ relays, for two values of $K\in \{ 10,20\}$ and the code rate $R=0.8$, corresponding to $N=13$ and $25$ coded transmissions, respectively. The finite field is binary and the PERs of the links originating at $\SSS$ are fixed to $\eSR=0.1$ and $\eSD=0.8$ this time, but $\eRD$ varies from $0.1$ to $0.8$. As a reference, the decoding probability calculated based on the original bound \eqref{eq:prob_dec_L} is also shown for $L \leq 4$. It can be observed that the simplified bound is accurate for $L \geq 2$ if $\eRD=0.1$ for both values of $K$, and for $L \geq 4$ and $3$ if $\eRD=0.5$ for $K=10$ and $20$, respectively. This is in line with condition \eqref{eq:condition}, which predicts the same values if $\delta=15$. At the same time, when the PEP between the relays and the destination node becomes high, such as $0.8$, bound \eqref{eq:simple_bound} is tight only for a relatively large number of relays, $8$ and $7$ for $K=10$ and $20$, respectively, which is again in line with \eqref{eq:condition}. It should be noted, however, that the network configuration with $\eRD=0.8$, shown in Fig.~\ref{fig:res_Simple_bound}, requires either a large number of relays or a much lower PEP between $\SSS$ and $\DDD$ to provide reliable delivery, and in both cases bound \eqref{eq:simple_bound} becomes tight. Finally, condition \eqref{eq:condition} also predicts that the simplified bound becomes tighter as $K$ increases for a given $L$, which is evident from Fig.~\ref{fig:res_Simple_bound}.

\subsection{Throughput Analysis\label{sec:Throughput}} 
\begin{figure}[t]
\subfloat[$\epsilon_\mathrm{SR}=0.1$]{\label{fig:res_TP1_1}
	\includegraphics[width=1\columnwidth]{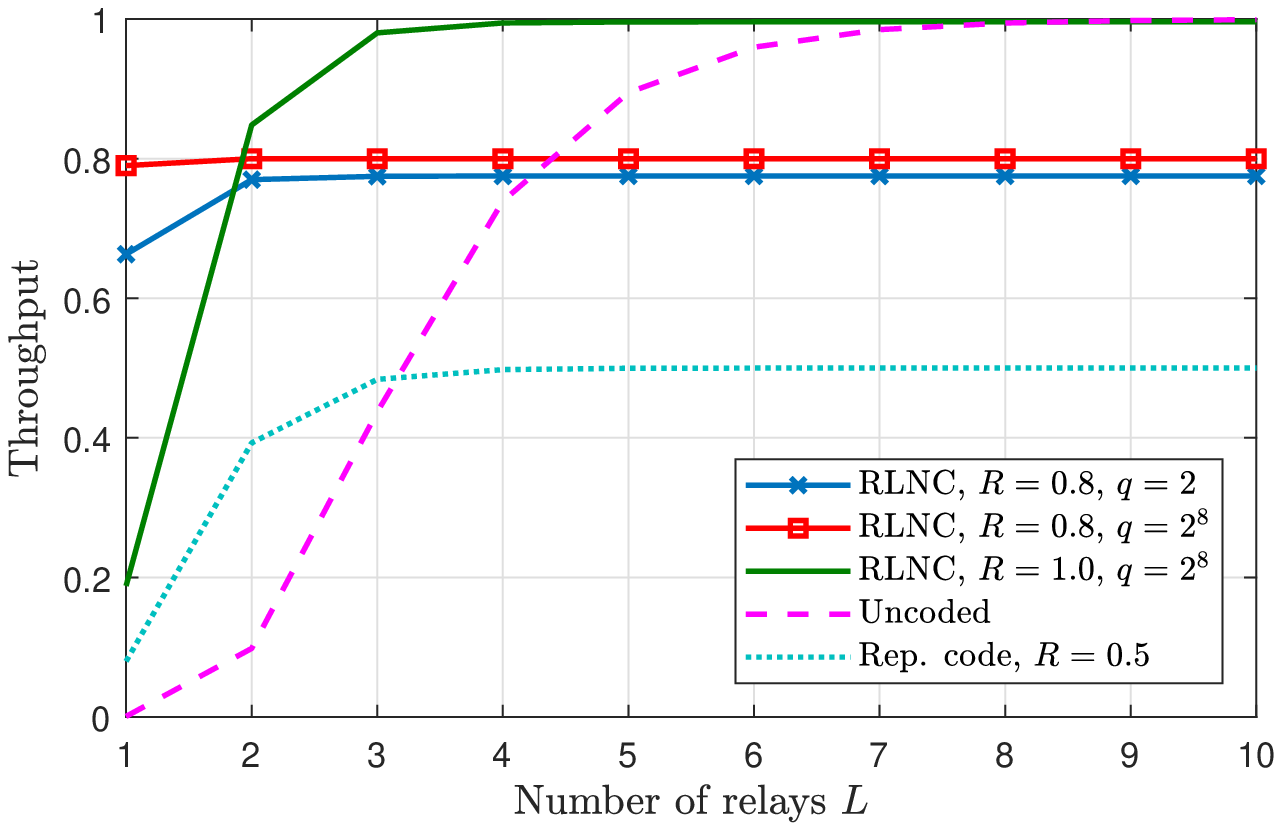}
}\\
\subfloat[$\epsilon_\mathrm{SR}=0.5$]{\label{fig:res_TP1_2}
	\includegraphics[width=1\columnwidth]{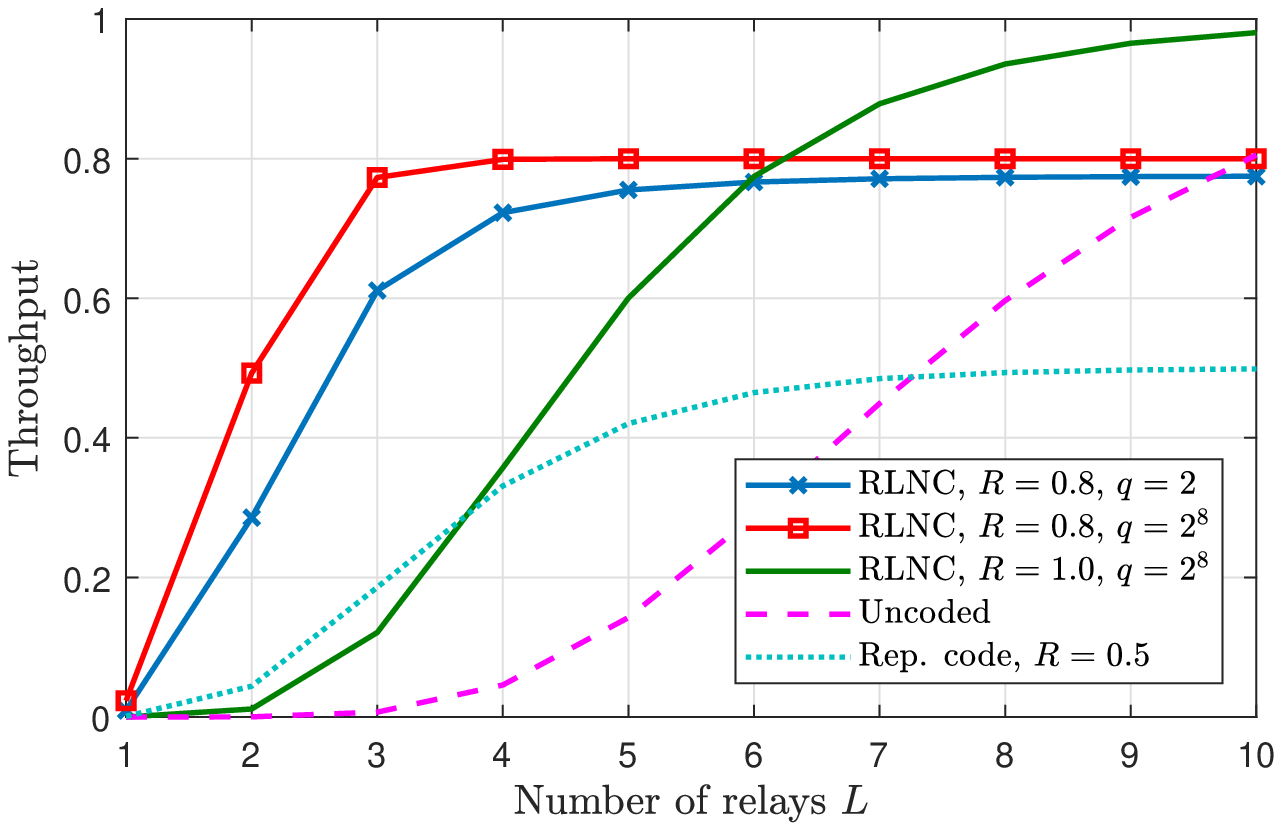}
}\\
\caption{Throughput of the RLNC-based scheme with $R \in \{0.8,1.0\}$, the uncoded system and the repetition code with $R=0.5$ for $K = 20$, $\eRD=0.3$, $\eSD=0.8$ and $\eSR=0.1$ (a) and $\eSR=0.5$ (b).}
\label{fig:res_TP1}
\end{figure}

Motivated by the results so far, we now employ the simplified bound \eqref{eq:simple_bound} to analyse system throughput and compare it with other coding and relaying approaches. To this end, we consider the decoding event as the successful delivery of a message of $K$ packets, given $N$ transmissions, and therefore define the Application-Layer throughput based on the decoding probability and the code rate \eqref{eq:code_rate} as follows:
\begin{equation}
	T_\mathrm{RLNC} = P_\RRR^{(L)} R.\label{eq:TP}
\end{equation}
We first compare the throughput of the proposed RLNC scheme with that of an uncoded system ($R=1$), in which the source packets are transmitted as they are. In addition, we consider a rate-$0.5$ repetition code, whereby each source packet is transmitted twice, from both the source and each relay node. The throughput in the latter case can be calculated as follows:
\begin{equation}
	T_\mathrm{Rep} = \left[ 1 - \left( \eSD \left[ (1-\eSR)\eRD+\eSR \right]^L \right)^{1/R} \right]^K R, \label{eq:TP_rep}
\end{equation}
which, by letting $R=1$, characterises the throughput of the uncoded system. Fig.~\ref{fig:res_TP1} illustrates the results for $K = 20$, $\eRD=0.3$, $\eSD=0.8$ and $\eSR \in \{0.1,0.5\}$, for different code rates $R \in \{0.8,1.0\}$ and finite field sizes  for the RLNC scheme. Based on condition \eqref{eq:condition} and the selected parameters, the simplified bound \eqref{eq:simple_bound} is expected to produce values very close to the original bound \eqref{eq:prob_dec_L} for $L \geq 2$. In the case when $\eSR=0.1$, it can be seen from Fig.~\ref{fig:res_TP1} that the RLNC scheme with $R=0.8$ significantly outperforms the system without coding for $L \leq 4$. The RLNC scheme with $R=1.0$ and $q=2^8$ achieves the maximum throughput when the number of relays is $5$, compared to $9$ for the uncoded system. In the case when $\eSR=0.5$, even the binary RLNC scheme with $R=0.8$ can achieve the same throughput as the uncoded system with $10$ relays, but having just $L=3$ relays. It can also be observed that with the larger PEP between the source and relay nodes, having a lower code rate is more beneficial if the number of relays is small. As regards the repetition code, it is clearly inferior to RLNC.

\begin{figure}[t]
\subfloat[$\epsilon_\mathrm{SR}=0.1$]{\label{fig:res_TP2_1}
	\includegraphics[width=1\columnwidth]{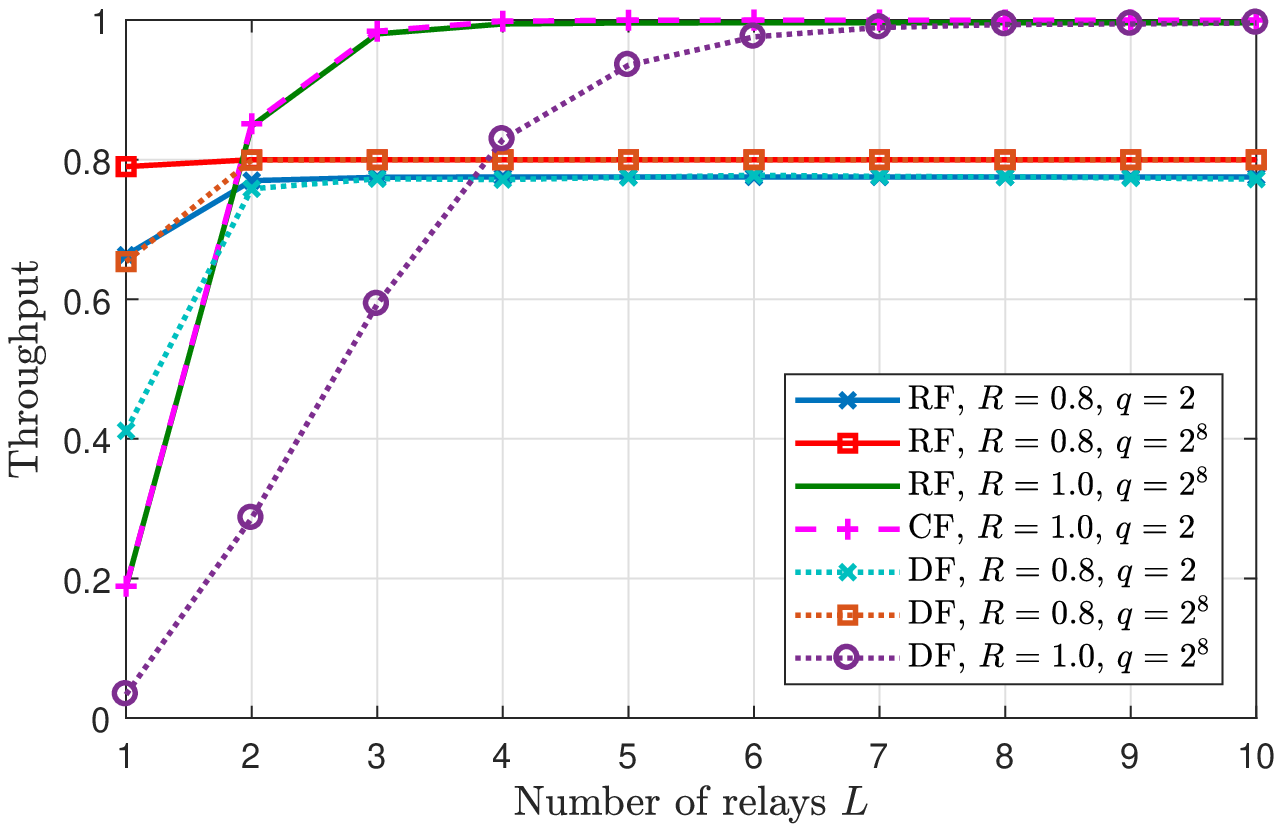}
}\\
\subfloat[$\epsilon_\mathrm{SR}=0.5$]{\label{fig:res_TP2_2}
	\includegraphics[width=1\columnwidth]{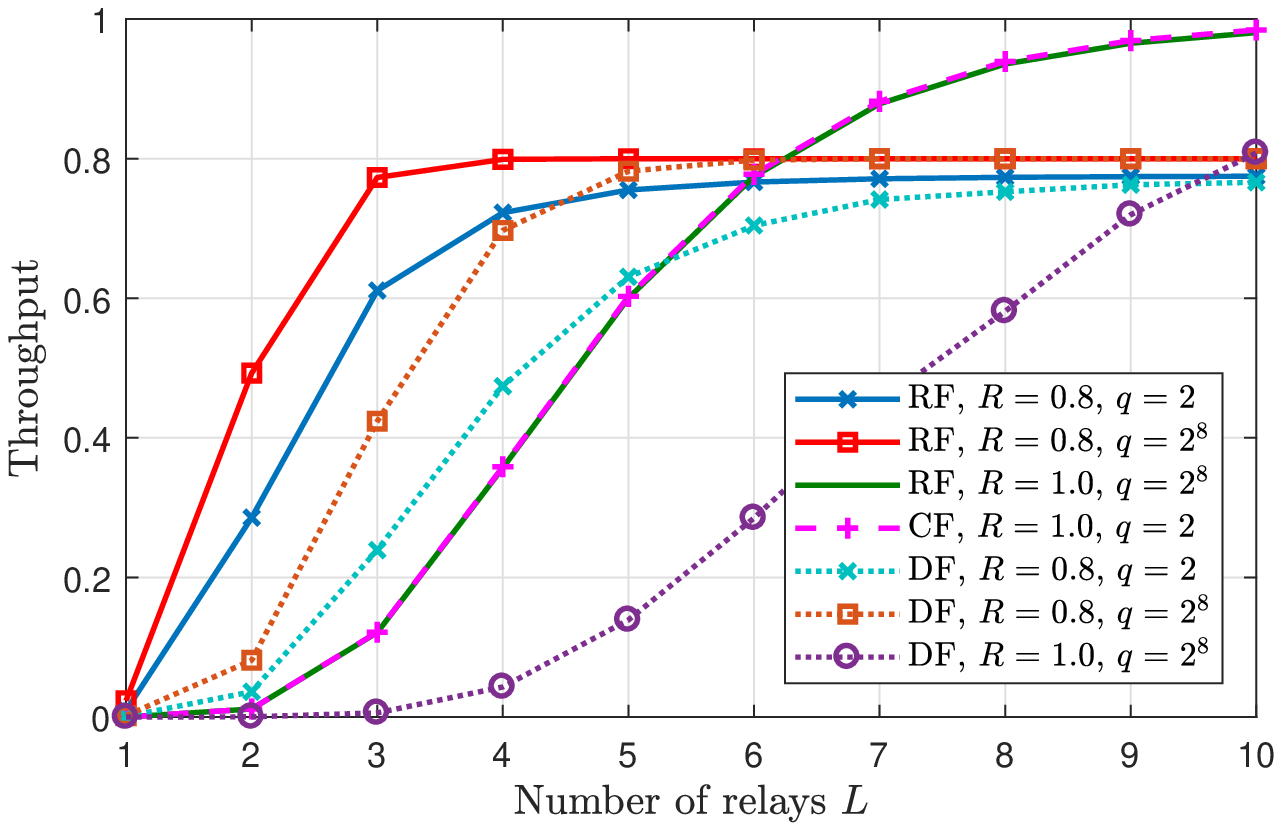}
}\\
\caption{Throughput comparison of different RLNC-based relaying strategies - RF, CF and DF - for $K = 20$, $\eRD=0.3$, $\eSD=0.8$ and $\eSR=0.1$ (a) and $\eSR=0.5$ (b).}
\label{fig:res_TP2}
\end{figure}

We now compare the throughput of the proposed scheme with other RLNC-based coding and relaying strategies. We consider three schemes: RF, analysed in this work, DF, briefly described in Section~\ref{sec:Introduction} and analysed in \cite{Khan2015,Tsimbalo2016a} for a single-relay network, and the RF strategy in which coding is performed only at the relay nodes, which will be referred to as \emph{Code-and-Forward (CF)}. The decoding probability of the CF scheme can be approximated by the same simplified bound \eqref{eq:simple_bound} as for the RF scheme by letting $N_\SSS = K$ and assuming no coding at $\SSS$, thus resulting in the following bound:
\begin{equation}
	P_\RRR^{(L)} \leq (1-\tilde{\epsilon})^K. \label{eq:bound_CF}	
\end{equation} 
In other words, decoding is successful if the union of the packets received by all relays and the destination node from $\SSS$ has size $K$. This bound does not depend on the finite field size and the number of coded transmissions of the relay nodes $N_\RRR$. Therefore, it will be assumed that the code rate of the CF scheme is equal to $1.0$ and the code is binary. As regards the DF scheme, due to the absence of the analytical results in the literature, the throughput of this scheme will be evaluated via simulation. Fig.~\ref{fig:res_TP2} illustrates the comparison between the three relaying strategies, for $K = 20$, $\eRD=0.3$, $\eSD=0.8$, $\eSR \in \{0.1,0.5\}$ and different code rates and finite field sizes where appropriate. It can be observed that the RF scheme provides a higher throughput than the DF strategy, especially with the larger $\eSR$ and the higher code rate. Indeed, for such scenarios, the relays operating under the DF scheme are not likely to decode and therefore will simply retransmit a fraction of packets transmitted by the source. By contrast, the relays operating under the RF scheme will always transmit $N_\RRR$ coded packets, thus increasing the probability of collecting $K$ linearly independent vectors of coding coefficients at $\DDD$. In other words, it is more beneficial \emph{not} to perform decoding at the relay nodes, but instead simply recode received packets. This is in line with the theoretical results reported in \cite{Lun2004, Lun2005} for classic RLNC, on which the RF scheme is based.

Compared with the CF strategy, the RF scheme employing binary codes is generally inferior for $\eSR=0.1$, while the non-binary RF scheme with $R=1.0$ provides the same throughput. Clearly, when packet loss between the source and relays is small, coding at $\SSS$ is redundant.  However, when $\eSR=0.5$, the RF scheme with $R=0.8$ becomes advantageous over the CF approach for $L \leq 6$. In other words, coding at $\SSS$ is beneficial if the PEP between the source and relays is high, unless the number of relays is sufficiently large to mitigate the packet loss.

Table~\ref{tab:summary_results} summarises the throughput results for all schemes considered in this section and highlights the best throughputs in bold.

\begin{table}
\setlength\tabcolsep{3.0pt}
\renewcommand*{\arraystretch}{1.2}
\scriptsize
\begin{centering}
\caption{Summary of throughput comparison results between various RLNC schemes,
the uncoded system and the repetition code, for $L=3$ and $8$ relays, $K=20$, $\epsilon_{\mathrm{RD}}=0.3$,
$\epsilon_{\mathrm{SD}}=0.8$ and different values of $\eSD$. Best throughputs are highlighted in bold. }
\label{tab:summary_results}
\begin{tabular}{|l|c|c|c|c|}
\hline 
\multirow{3}{*}{Coding scheme} & \multicolumn{4}{c|}{Throughput}\tabularnewline
\cline{2-5} 
 & \multicolumn{2}{c|}{$L=3$} & \multicolumn{2}{c|}{$L=8$}\tabularnewline
\cline{2-5} 
 & $\epsilon_{\mathrm{SR}}=0.1$ & $\epsilon_{\mathrm{SR}}=0.5$ & $\epsilon_{\mathrm{SR}}=0.1$ & $\epsilon_{\mathrm{SR}}=0.5$\tabularnewline
\hline 
\hline 
Uncoded & $0.44$ & $0.01$ & $\boldsymbol{0.99}$ & $0.60$\tabularnewline
\hline 
Rep. code, $R=0.5$ & $0.48$ & $0.19$ & $0.50$ & $0.49$\tabularnewline
\hline 
RLNC RF, $R=0.8$, $q=2$ & $0.77$ & $0.61$ & $0.78$ & $0.77$\tabularnewline
\hline 
RLNC RF, $R=0.8$, $q=2^{8}$ & $0.80$ & $\boldsymbol{0.77}$ & $0.80$ & $0.80$\tabularnewline
\hline 
RLNC RF, $R=1.0$, $q=2^{8}$ & $\boldsymbol{0.98}$ & $0.12$ & $\boldsymbol{1.00}$ & $\boldsymbol{0.94}$\tabularnewline
\hline 
RLNC CF, $R=1.0$, $q=2$ & $\boldsymbol{0.98}$ & $0.12$ & $\boldsymbol{1.00}$ & $\boldsymbol{0.94}$\tabularnewline
\hline 
RLNC DF, $R=0.8$, $q=2$ & $0.77$ & $0.24$ & $0.78$ & $0.75$\tabularnewline
\hline 
RLNC DF, $R=0.8$, $q=2^{8}$ & $0.80$ & $0.42$ & $0.80$ & $0.80$\tabularnewline
\hline 
RLNC DF, $R=1.0$, $q=2^{8}$ & $0.59$ & $0.01$ & $0.99$ & $0.58$\tabularnewline
\hline 
\end{tabular}
\par\end{centering}
\end{table}

\section{Conclusions and Future Work\label{sec:Conclusions}}
In this paper, we have addressed the theoretical performance analysis of a single-source relay-assisted network operating under RLNC. We employed the classical network coding scheme, in which the relay nodes simply recode packets they receive, without resorting to decoding. We proposed a novel bound \eqref{eq:prob_dec_L} for the decoding probability of a network with an arbitrary number of relays, which becomes exact for a single-relay network. Furthermore, we proposed a simple, scalable approximation \eqref{eq:simple_bound} of bound \eqref{eq:prob_dec_L} valid for a sufficiently large number of relays.

The theoretical results were verified via Monte Carlo simulation for various network and code parameters. The proposed bound was shown to be very tight, with the accuracy increasing as the number of relays or the source message size grows. In the case of a single relay, it was demonstrated that the theoretical results perfectly match the simulated ones. We also showed that for a sufficiently large number of relays, the approximated version of the bound is equally suitable and is very tight to the simulated performance.

To show the advantages of the considered RLNC-based coding scheme, we performed a throughput comparison with an uncoded case, as well as with a repetition code. It was demonstrated that the repetition code is inferior to RLNC, while the uncoded scheme requires a much larger number of relays to achieve the same reliability. Finally, we verified the necessity of coding at the source node and justified the proposed relaying strategy. In both cases, the benefit of the proposed scheme is especially large if the packet loss is high. The throughput results are summarised in Table~\ref{tab:summary_results}.

The novel theoretical framework proposed in this paper can be used to predict the performance of relay networks and to select optimum network and code parameters without resorting to complex simulations. In particular, the proposed simplified bound can be employed for the analysis of other performance metrics, such as delay and energy efficiency. In the future, it is planned to extend the framework to networks with multiple sources and destinations, as well as to consider other coding strategies, such as systematic and sparse RLNC.

\section*{Appendix A}

\section*{Proof of Lemma~\ref{lemma:prob_dest_full_rank}}
Consider the second term under the summation in \eqref{eq:prob_dest_full_rank}, the probability of an $(r+m_\DDD) \times K$ matrix having rank $K$. This matrix can be viewed as two vertically stacked matrices with $r$ and $m_\DDD$ rows. Hence, applying \eqref{eq:prob_rank_vert} yields:
\begin{equation}
	\PP(r+m_\DDD, K) = \sum_{i=i_{\min}}^{i_{\max}} \PP_i(m_\DDD,K) \PP(r, K-i), \label{eq:lemma2_proof_1}
\end{equation}
where $i_{\min} = \max(0,K-r)$ and $i_{\max} = \min(m_\DDD,K)$. Substituting \eqref{eq:lemma2_proof_1} and \eqref{eq:lemma_bound} into \eqref{eq:prob_dest_full_rank} results in the following bound:
\begin{equation}
	\mathrm{Pr}\left[ X \right] \leq \hspace{-2mm}\sum_{r=r_{\min}}^{r_{\max}}\hspace{-2mm} \PP_r(m', m-\mRD) %
							      \hspace{-1mm}\sum_{i=i_{\min}}^{i_{\max}}\hspace{-1mm} \PP_i(m_\DDD,K) \PP(r, K-i). \label{eq:lemma2_proof_2}
\end{equation}
Changing the order of summation and updating the starting indices to \mbox{$i_{\min} = \max(0,K-m+\mRD)$}, \mbox{$r_{\min} = \max(0,K-i)$}, \eqref{eq:lemma2_proof_2} can be rewritten as follows:
\begin{equation}
	\mathrm{Pr}\left[ X \right] \leq \hspace{-2mm}\sum_{i=i_{\min}}^{i_{\max}}\hspace{-2mm} \PP_i(m_\DDD,K) %
							      \hspace{-1mm}\sum_{r=r_{\min}}^{r_{\max}}\hspace{-1mm} \PP_r(m', m-\mRD) \PP(r, K-i). \label{eq:lemma2_proof_3}
\end{equation}
Consider the inner sum in \eqref{eq:lemma2_proof_3} and let, for compactness, \mbox{$m''=m-\mRD$} and \mbox{$K'=K-i$}. Assume, without loss of generality, that \mbox{$m''\leq m'$}. The sum can be rewritten as follows:
\begin{eqnarray}
	&   &\hspace{-5mm}\sum_{r=r_{\min}}^{r_{\max}} \PP_r(m', m'') \PP(r, K') = \PP(m'', K')  \PP(m', K') \nonumber\\		
	&	&\hspace{-5mm}\cdot \sum_{r = K'}^{m''} \frac{q^{K'(m' + m'')}}{q^{rK' + m'm''}} \prod_{l=K'}^{r-1} \left(\frac{q^{m''}-q^{l}}{q^r-q^l}\right) \left(q^{m'}-q^{l}\right).\label{eq:identity}
\end{eqnarray}
The last sum can be shown to be equal to $1$ by changing the summation index to $r' = r - K'$ and realising that the summation operand is equal to \mbox{$\PP_{r'}\left( m'-K', m''-K' \right)$}. As a result, returning to the original notation, \eqref{eq:lemma2_proof_3} can be rewritten as follows:
\begin{equation}
	\mathrm{Pr}\left[ X \right] \leq \hspace{-2mm}\sum_{i=i_{\min}}^{i_{\max}}\hspace{-2mm} \PP_i(m_\DDD,K) \PP(m-\mRD, K-i) \PP(m', K-i). \label{eq:lemma2_proof_4}
\end{equation}
It can be observed that the right-hand side of \eqref{eq:lemma2_proof_4} is identical to that of \eqref{eq:prob_two_mat_fr}, where instead of $m_1$, $m_2$ and $m_{12}$ one uses \mbox{$m-\mRD+m_\DDD$}, \mbox{$m'+m_\DDD$} and \mbox{$m_\DDD$}, respectively.
$\hfill\blacksquare$

\bibliographystyle{ieeetr}
\bibliography{Relay_RLNC}

\end{document}